\documentclass[english,12pt,aps,nofootinbib]{revtex4-1}
\usepackage{lmodern}
\usepackage[T1]{fontenc}
\usepackage[latin9]{inputenc}
\usepackage{color}
\usepackage{babel}
\usepackage{array}
\usepackage{booktabs}
\usepackage{textcomp}
\usepackage{amsmath}
\usepackage{graphicx}
\usepackage{subscript}
\usepackage[unicode=true,pdfusetitle,
 bookmarks=true,bookmarksnumbered=false,bookmarksopen=false,
 breaklinks=false,pdfborder={0 0 1},backref=false,colorlinks=true]
 {hyperref}
\hypersetup{
 linkcolor=magenta, citecolor=blue, urlcolor=blue, pdfstartview=XYZ, plainpages=false, pdfpagelabels}

\makeatletter

\providecommand{\tabularnewline}{\\}

\PassOptionsToPackage{square,comma}{natbib}
\RequirePackage{natbib}
\makeatother

\begin{document}
\title{Enhanced adsorption of Xe and Kr on boron doped graphene
sheet decorated with transition metals (Fe, Ni, Cu and Zn)}
\author{P. Anees}
\email{anees@igcar.gov.in}

\affiliation{\textsuperscript{}Materials Physics Division, Indira Gandhi Centre
for Atomic Research, Kalpakkam 603 102, TamilNadu, India}
\begin{abstract}
Efficient adsorption and segregation of Xe \& Kr gases is of high
importance in commercial as well as nuclear industries{\normalsize{}.
Systematic }\textit{\normalsize{}ab initio}{\normalsize{} calculations
reveal that transition-metal (TM) decorated boron-doped graphene (BDG-TM)
sheet can act as an efficient substrate for adsorptive capture of
Xe \& Kr (adatoms). Substantial enhancement in the adsorption energy
}$(E_{ads})$ {\normalsize{}is obtained on BDG-TM substrates} and
it varies as BDG-Cu > BDG-Ni > BDG-Fe > BDG-Zn. The improvement is
approximately four times that of the pristine BDG and twice that of
the conventional metallic substrates. Bader charge analysis and charge
density difference maps envisage that, the TM-decoration alters the
charge distribution at substrate-adatom interface, which in-turn brings
a considerable change in the polarization of adatom, leading to significant
improvement in the $E_{ads}$. The change in polarization of adatoms
is interlinked with charge transfer process and it has been gauged
by computing their effective charges, which follows the same sequence
of $E_{ads}$ and hence corroborated each other. Later, the partial
density of states analysis shows a splitting and significant interaction
of Xe-\textit{p} with TM-\textit{d} orbitals near the Fermi level
of Fe, Ni and Cu decorated systems, unveiling a strong adsorption.
Further, the effect of clustering and dispersion of Cu atoms on $E_{ads}$
are analyzed using a first principle based genetic algorithm, which
reveals that clustering of Cu atoms deteriorate the $E_{ads}$ of
Xe \& Kr. Thus for experimental realization, BDG sheet with uniformly
dispersed fine Cu particles is proposed as a substrate.
\end{abstract}
\maketitle

\section{Introduction}

Xenon (Xe) and Krypton (Kr) are the primary components of gaseous
fission products, known as fission gases (FGs). Radioactive FGs  infiltrate
to atmosphere, during the reprocessing of spent nuclear fuels, clad
rupturing under operating conditions, nuclear detonations and the
production of medical isotopes \citep{soelberg_radioactive_2013,REST2019310,hoffman_medical_2018}.
Proper monitoring, segregation and disposal of such radioactive gases
are essential to prevent the atmospheric contamination. On the other
hand, Xe has several industrial applications as well; it's being widely
used in light sources, in the field of medical anesthesia, protein
crystallography and satellite propulsion system etc \citep{Dmochowski2009}.
Hence an efficient capture and segregation of Xe is a mandate of both
nuclear and commercial industries. 

Traditionally, adsorptive methods have been widely used for capture
and segregation of rare gases. Adsorption of rare gases on surfaces
are classical problems in surface science \citep{bruch1997physical}.
Metal surfaces are the conventional substrates for physical adsorption
of Xe \& Kr \citep{doi:10.1063/1.446815,VIDALI1991135,PhysRevLett.90.066104,PhysRevB.68.045406,PhysRevB.77.045401,Chen_2012,PhysRevB.91.195405}.
On metallic substrates, the adsorption energies ($E_{ads}$) of rare
gases ranges from $\approx$ -10 to - 400 meV \citep{PhysRevB.77.045401};
for example, the $E_{ads}$ of He/Pd(110), Kr/Pd(100), Xe/Pd(111)
and Xe/Ni(111) systems are -10, -200, -320 to -360 and -401 meV, respectively
\citep{doi:10.1063/1.446815,VIDALI1991135,PhysRevLett.90.066104,PhysRevB.68.045406,PhysRevB.77.045401,PhysRevB.91.195405}.
Zeolites and activated carbons were also used as substrates for adsorption
and segregation of Xe \& Kr \citep{bazan_adsorption_2011,thallapally_facile_2012}.
Later, metal organic frame works (MOFs) \citep{BANERJEE2018466,C7TA11321H,C8TA02091D,C3SC52348A,doi:10.1021/ar5003126,doi:10.1021/jp501495f},
organic porous cage molecules \citep{chen_separation_2014} and carbon
nano cages \citep{Rahimi_2018} were adopted for Xe \& Kr adsorption.
Banerjee \textit{et al }\citep{banerjee_metalorganic_2016} reported
a calcium-based nanoporous MOF (SBMOF-1) with the highest Xe adsorption
energy of $\approx$ -42 kJ/mol (-435 meV). 

In recent years, search for effective alternative substrates are in
top gear. Properties such as, large surface area, functionality, light
weight and stability of graphene based materials enables it a promising
candidate for gas adsorption and storage applications \citep{GADIPELLI20151}.
Graphite surfaces were used as substrates to adsorb the Xe atoms \citep{CHEN2003403,PUSSI2004157,PhysRevB.76.085301}.
\textit{Ab initio} studies revealed the $E_{ads}$ of Xe on hollow
sites of graphite are -204 \citep{CHEN2003403} and -168 \citep{PhysRevB.76.085301}
meV. Sheng \textit{et al} \citep{doi:10.1021/jp907861c} computed
the adsorption energies of Xe on graphene using \textit{ab initio}
MP2 calculation. In this study, they modeled graphene as polycyclic
aromatic hydrocarbons (coronene). The Xe atom prefers to bind at the
hollow site of coronene with $E_{ads}$ of -142.9 meV. Later, Ambrosetti\textit{
et al} \citep{doi:10.1021/jp110669p} made a comprehensive analysis
of adsorption of rare gases on graphene and graphite using van der
Waals (vdW) corrected DFT calculations and predicted the $E_{ads}$
of Xe on graphene as -171.2 meV. Subsequently, hetero-atom doping
in graphene sheet has been tried  to enhance interaction between the
rare gases and sheet substrate \citep{VAZHAPPILLY2017174}. In this
attempt, authors employed both n \& p type dopants to refine the adsorption
parameters of Xe \& Kr.  Among the different dopants, Be-doped graphene
sheet yield marginal enhancement in $E_{ads}$ of Xe (-196.70 meV)
and Kr (-140.63 meV) \citep{VAZHAPPILLY2017174}. The hetero-atom
doping brings about 11.41 \% enhancement in adsorption energy of Xe
with respect to pristine graphene. Later, the authors extended their
studies to graphyne and graphdiyne systems with various chemical dopants
\citep{VAZHAPPILLY2020100738}. Aluminum (Al) doped graphyne and graphdiyne
shows an improvement in the $E_{ads}$ of Xe \& Kr with respect to
pristine. The highest $E_{ads}$ obtained is -348.5 meV for Xe adsorbed
on top site of Al-graphdiyne. Very recently, pristine and doped phosphorene
sheets have been employed to adsorb the rare gas atoms \citep{HU2020144326}.
The $E_{ads}$ on pristine phosphorene sheets are comparable to that
of graphene, while the Li doped phosphorene sheet shows an enhancement
in the Xe (-294.7 meV) and Kr (-237.4 meV) interaction with substrate.

From the aforementioned discussions, it can be seen that the upper
bound of $E_{ads}$ of Xe is in the range of -400 to -450 meV. The
magnitude of $E_{ads}$ indicates a weak adsorption, and hence further
enhancement in adsorption energy is essential for efficient adsorptive
capture and segregation of Xe \& Kr. This motivated us to look for
alternate substrates, which can emanate a substantial enhancement
in the adsorption energy. Transition metals (TMs) decoration on carbon
nano structures has resulted in considerable improvement in the adsorption
properties of gases and molecules \citep{NACHIMUTHU2014132,JUNGSUTTIWONG2016140,CORTESARRIAGADA2018227}.
Motivated from this fact, a new strategy is adopted; where TMs (Fe,
Ni, Cu and Zn) decorated boron-doped graphene (BDG) sheets are employed
as substrates for efficient adsorption and separation of Xe \& Kr.
The aforementioned TMs are relatively cheap and earth abundant in
comparison to noble metals. Here, instead of pristine graphene, BDG
sheets are used for TM-decoration due to the following reasons. Earlier
studies reported that boron-doped graphene sheets can adsorb hydrogen
more strongly than pristine graphene due to strong boron binding sites
\citep{SANKARAN2008393,wu_dft_2011}. Also, boron doping forbids the
clustering and agglomeration of metal atoms \citep{BEHESHTI20111561},
which can affect the adsorption properties. More importantly, boron-doped
graphene sheets have been successfully synthesized experimentally
\citep{SHIRASAKI20001461}. Henceforth, in this work, systematic \textit{ab
initio} density functional theory (DFT) calculations have been performed
to analyze the adsorption parameters and mechanism of Xe \& Kr on
these newly proposed substrates. A substantial improvement in the
$E_{ads}$ is obtained, and its underlying mechanism is unveiled from
the detailed Bader charge, induced dipole moment, charge density difference
maps and partial density of states analysis of each BDG-TM-adatom
systems. Further, the effects of nano-clustering and dispersion of
TMs on adsorption parameters are analyzed. At end, an ATLAS of $E_{ads}$
of Xe is made and compared with prior studies to highlight the outcome
of the present study. 

\section{Computational methods\label{sec:Computational-methods}}

Spin polarized \textit{ab initio} DFT calculations are done using
Vienna \textit{ab initio} simulation package (VASP) \citep{PhysRevB.54.11169}.
To ensure the accuracy and efficiency of calculations, projector augmented
wave (PAW) pseudo potential is adapted. Generalized gradient approximation
(GGA) of Perdew--Burke--Ernzerhof (PBE) is used to compute the exchange
correlation functional \citep{PhysRevLett.77.3865}. The size of the
plane wave basis set is truncated using a kinetic energy cut-off of
500 eV. A Monkhorst \citep{PhysRevB.13.5188} Pack grid of size 5\texttimes 5\texttimes 1
is used to sample over the Brillouin zone. All the geometries were
optimized with a force convergence of the order of 10\textsuperscript{-2}
eV/Å. In geometry optimization, the Xe/Kr were placed on a pre-converged
BDG-TM sheets and their coordinates and cell shapes were relaxed by
keeping the coordinates of the substrate atoms fixed. This particular
choice of geometry optimization scheme is motivated by references
\citep{doi:10.1021/jp110669p,PhysRevB.76.085301,doi:10.1021/acs.jpcc.6b02782,VAZHAPPILLY2017174}.
To compute the PDOS and charge density, a dense k-point grid of size
15x15x3 is used. Graphene sheet of supercell size 6x6x1 (72 atoms)
has been selected for boron doping and then metal-decoration. A vacuum
separation of 15 Å is provided along \textit{c}-axis to avoid the
un-physical interactions between periodic images. The substrate-adatom
van der Waals (vdW) interaction is incorporated using the scheme of
Tkatchenko \textit{et al.} (TS-SCS) \citep{PhysRevLett.108.236402},
which has been successfully used to model the binding of transition
metals and their clusters on graphene \citep{PhysRevB.95.235422},
noble gas solids \& layered materials \citep{al-saidi_assessment_2012},
water \& ethanol adsorption on different TM surfaces \citep{doi:10.1021/acs.jpcc.7b09749}
etc. In TS-SCS method, the electrostatic screening effect is computed
in a self-consistent manner. Also, it incorporates the global interactions
of fluctuating dipoles, that enables us to treat system as a whole,
thus going beyond the conventional pair wise methods. 

\section{Results and Discussions \label{sec:Results-and-Discussions}}

\subsection{Transition-metal (TM)-decoration on Boron-doped graphene (BDG) \label{subsec:BDG-TMs-prop}}

\begin{figure*}
\begin{centering}
\includegraphics[scale=1.5]{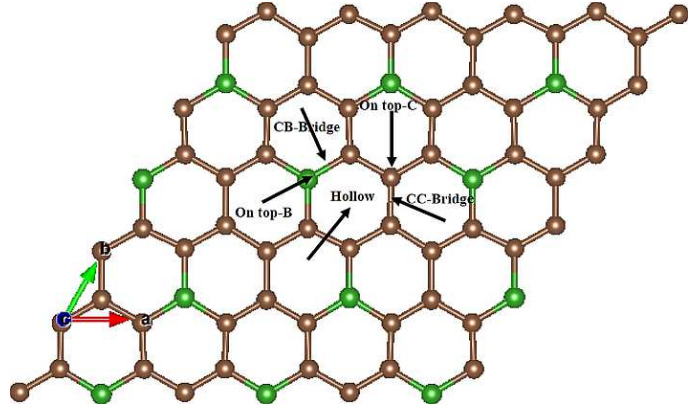}
\par\end{centering}
\caption{\label{fig:sheet-BDG}Optimized geometry of boron-doped graphene (BDG)
sheet, brown and green colour spheres represent the C and B atoms.
The possible binding sites such as hollow (H), C-C bridge, C-B bridge,
on top of C and on top of B atoms are marked. }
\end{figure*}
 In-order to construct the boron-doped graphene (BDG), firstly, a
6x6x1 supercell of graphene sheet is made. Earlier experimental \citep{PhysRevB.54.144}
and theoretical \citep{NACHIMUTHU2014132,JUNGSUTTIWONG2016140} studies
show that an ordered C\textsubscript{5}B layer is the energetically
favourable configuration for BDG sheet. Hence, B atoms are substitutionally
doped at each hexagonal ring of the pre-optimized graphene sheet,
which generates an ordered layer structure with stoichiometry C\textsubscript{60}B\textsubscript{12},
which leads to $\approx$16 \% doping. Such heavily doped BDG sheets
were previously employed as an efficient substrate for electro-catalytic
and hydrogen storage applications \citep{C5TA10599D,NACHIMUTHU2014132}.
Figure \ref{fig:sheet-BDG} shows the optimized structure of BDG sheet,
which retains the planar geometry of pristine graphene. The energetics
of B doping is analyzed from the cohesive energy $(E_{coh})$, calculated
as\vspace{-0.3cm}
\begin{align}
E_{coh} & =(E_{total}(BDG)-n_{i}E_{i})/N
\end{align}
Where the $E_{total}(BDG)$, $n_{i}$, $E_{i}$ and $N$ are the total
energy of BDG sheet, number of dopant (B) atoms, energy per atom of
dopant and total number of atoms in supercell, respectively. The $E_{coh}=$
-7.493 eV/atom; the negative cohesive energy indicates the energetic
stability of BDG system. Further, the formation energy of BDG have
been computed as follows,

\vspace{-0.75cm}
\begin{align}
E_{form} & =\{E_{total}(BDG)-E_{total}(Gr)-n_{i}(\mu_{B}-\mu_{c})\}/n_{i}
\end{align}
Where, $E_{total}(BDG)$, $E_{total}(Gr)$, $n_{i}$, $\mu_{B}$ and
$\mu_{c}$ are the total energy of BDG, total energy of pristine graphene,
number of dopant (B) atoms, chemical potentials of boron and carbon
atoms, respectively. The $\mu_{B}$ and $\mu_{c}$ are taken as the
binding energy per atom of alpha-boron and graphene sheet, respectively.
The computed formation energy per B atom is 1.448 eV/B-atom, which
is matching with previous prediction \citep{doi:10.1063/1.5018065}. 

\begin{figure*}
\centering{}\includegraphics[scale=0.45]{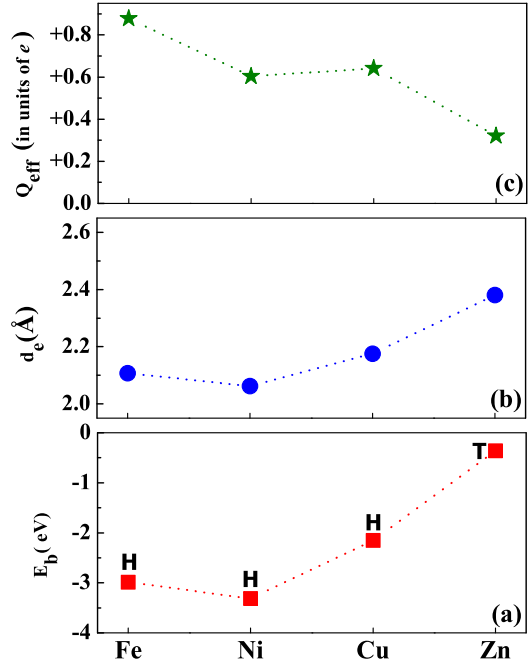}\caption{\label{fig:BDG-NDG-TM-ads-prop}(a) Binding energy ($E_{b}$), (b)
equilibrium distance ($d_{e}$) between TM atom \& BDG sheet, (c)
effective charge ($Q_{eff}$) on TM atom (in units of \textit{e} ),
for TM decorated BDG sheet. The preferred adsorption sites (H \& T)
for each TMs are indicated.}
\end{figure*}

In BDG, the possible binding sites for TM atoms are labeled as shown
in figure \ref{fig:sheet-BDG}. Optimized geometries of BDG-TMs shows
that Fe, Ni and Cu atoms prefers to occupy the hollow position, while
Zn atom sits on top of B atom (Figure S1 -\textit{ SI}), it's in agreement
with earlier reports \citep{NACHIMUTHU2014132,JUNGSUTTIWONG2016140}.
The binding energy ($E_{b}$) of TM atoms (TMs) on BDG sheet is computed
as,\vspace{-0.25cm}
\begin{align}
E_{b} & =E_{total}-E_{sheet}-E_{TM}
\end{align}
Where, $E_{total}$, $E_{sheet}$ and $E_{TM}$ are the total energy
of BDG sheet with TM atom, BDG sheet, and TM atoms in gas phase, respectively.
Figure \ref{fig:BDG-NDG-TM-ads-prop}a shows the binding parameters
of TMs on BDG sheet. The calculated binding energy ($E_{b})$ of Fe,
Ni, Cu and Zn are -2.988, -3.312, -2.150 and -0.363 eV, respectively.
The $E_{b}$ of TMs on BDG is higher than that of pristine graphene
\citep{NAKADA201113}, signifying that boron doping in graphene enhances
the binding of TM atoms. For example, the $E_{b}$ of Ni on graphene
is -1.470 eV \citep{PhysRevB.95.235422} the same on BDG is -3.312
eV. This increase in $E_{b}$ helps to trap the TMs at binding sites
and hence prevents the clustering of TMs on BDG \citep{BEHESHTI20111561,NACHIMUTHU2014132}.
Boron is a p-type dopant which makes sheet more electron deficient
in comparison to pristine sheet. Hence the electron rich TM atom interact
strongly with electron deficient BDG sheet. The binding energy of
TM atom is expected to decrease as 3\textit{d} orbital populates \citep{Santos_2010}.
However, in the present study, the $E_{b}$ obtained for Ni is higher
than that of Fe, this is in line with the observation of Johll \textit{et
al }\citep{PhysRevB.79.245416}. This non-monotonic behaviour attributes
to the lower inter-configurational energy of Ni with respect to Fe.
That means the energy required to transfer unit charge from \textit{s}
orbital to \textit{d} (i.e. $3d^{n-2}4s^{2}\longrightarrow3d^{n-1}4s^{1})$
is lower for Ni in comparison with Fe, which results in strong binding
of Ni with sheet \citep{doi:10.1063/1.480546,PhysRevB.79.245416}.
In the case of Zn, the filled \textit{s }and \textit{d }orbitals leads
to a smaller binding energy. The equilibrium distance ($d_{e}$) between
the TM atom and the BDG sheet is shown in figure \ref{fig:BDG-NDG-TM-ads-prop}b.
The $d_{e}$ value of Fe and Ni are very close to each other. For
Cu and Zn the $d_{e}$ increases, attributing to their filled 3\textit{d}
orbitals and large size of 4\textit{s} orbitals \citep{Santos_2010}.
The higher $d_{e}$ value of Zn atom is corroborated with its low
binding energy. 

In-order to understand the nature of charge transfer upon TM binding,
the effective charges ($Q_{eff}$) on TMs are computed using Bader
charge analysis \citep{Tang_2009}. In Bader method, the total electronic
charge density on an atom is approximated to charge density encompassed
in a certain volume, known as Bader volume. The Bader charge ($Q_{badar}$)
on each atom is computed using the Henkelman's method \citep{Tang_2009}.
Thereafter, the $Q_{eff}$ is obtained as,
\begin{equation}
Q_{eff}=Z_{val}-Q_{badar}
\end{equation}
Where, $Z_{val}$ is the valency of atom. Figure \ref{fig:BDG-NDG-TM-ads-prop}c
shows the $Q_{eff}$ on TMs in electronic (\textit{e}) unit. The $Q_{eff}$
of all TM atoms are positive (Figure \ref{fig:BDG-NDG-TM-ads-prop}c),
indicating that the charge transfer is from TM atom to BDG sheet.
Since the single TM atom binds to the sheet, the charge transfer can
be explained based on their electronegativity values, which are 1.83,
1.91, 1.90 and 1.65, for Fe, Ni, Cu and Zn, respectively. The observed
$Q_{eff}$ values are in accordance with electronegativity data, except
for Zn. The charge transfer is highest for Fe, among the Cu and Ni,
the charge transfer is marginally higher for Cu, and it owes to its
slightly lower electronegativity with respect to Ni. For Zn, the completely
filled 3\textit{d }and 4\textit{s }orbital reduces the charge transfer
to BDG sheet. 

\subsection{Adsorption of Xe \& Kr on TM-decorated BDG sheet \label{subsec:BDG-TM-adatom-prop}}

In this section, the adsorption parameters of Xe \& Kr (adatoms) on
pristine graphene, BDG and TM decorated BDG (BDG-TM) substrates are
analyzed and compared. In pristine graphene, adatoms occupies the
H site (Figure S2-\textit{SI}). The adsorption energy ($E_{ads}$)
on different substrate is computed as,
\begin{align}
E_{ads} & =E_{total}-E_{substrate}-E_{a}
\end{align}
Where, $E_{total}$, $E_{substrate}$ and $E_{a}$ are the total energy
of substrate-adatom, substrate and adatoms in gas phase, respectively.
In proceeding section, $E_{ads}$ is expressed in meV to have a consistent
comparison with literature. Simple GGA (PBE) functional predicts a
negligibly small $E_{ads}$ (-9.0 meV) and a large adatom-sheet equilibrium
distance - $d_{e}^{a}$ (4.517 \AA). This poor binding in simple
GGA calculation envisages its inadequacy to explain the adsorption
processes of Xe \& Kr and it attributes to the non-accountability
of vdW interaction in these calculations\citep{doi:10.1021/jp110669p}.
The adatom and substrate interaction are non-covalent in nature and
primarily arising from the vdW interactions of separated fragments
of the system. Therefore the adsorption parameters of Xe \& Kr are
computed using vdW corrected GGA (PBE+vdW) calculations.

Table \ref{tab:ads-prop_Pristine_BDG-Gr} shows the adsorption parameters
of Xe \& Kr on graphene and BDG sheets. The $E_{ads}$ of Xe \& Kr
on graphene are -175 \& -133 meV, which shows an excellent agreement
with previous calculations (Table \ref{tab:ads-prop_Pristine_BDG-Gr}).
However, the PBE+vdW - $d_{e}^{a}$ values are over estimated with
respect to the earlier reports \citep{doi:10.1021/jp110669p,VAZHAPPILLY2017174,PhysRevB.76.125112}.
There is a marginal spread in the $d_{e}^{a}$ values among different
calculations. This discrepancy is due to the different choice of vdW
schemes in those studies, which has already been reported in references
\citep{PhysRevB.76.125112,doi:10.1021/jp110669p,SHEPARD201938}. Despite
the differences in $d_{e}^{a}$, the adsorption energetics and sites
are predicted consistently among different vdW correction schemes,
guaranteeing an accurate prediction of the adsorption energy. The
$E_{ads}$ of Xe \& Kr on BDG are -178 \& -137 meV, which is slightly
higher\footnote{In this manuscript, usage of higher/increase/enhancement in adsorption
energy ($E_{ads}$) means, the $E_{ads}$ becomes more negative}than that of the previous report \citep{VAZHAPPILLY2017174}. This
small disparity ascribes to the different choice of adsorption sites;
in their studies, $E_{ads}$ is computed on top site of dopant atom.
Whereas, in the present study, after geometry optimization, the adatoms
move from the initial H site into B site of BDG (Figure S2 -\textit{
SI}). Also, the authors used a single-atom doped graphene sheet (C\textsubscript{71}B\textsubscript{1})
as substrate, which is very dilute doping ($\approx$1 \%) with respect
to the BDG sheet (C\textsubscript{60}B\textsubscript{12}) employed
in the present study ($\approx$16 \%). This shows that the adsorption
capacity of sheet improves with the boron doping concentration. The
$E_{ads}$ of Xe is higher than that of Kr on graphene and BDG. This
difference in $E_{ads}$ is correlated to the difference in atomic
polarizability of Xe (3.99 x 10\textsuperscript{\textminus 24} cm\textsuperscript{3})
and Kr (2.46 x 10\textsuperscript{\textminus 24} cm\textsuperscript{3})
\citep{kittel2004introduction,PhysRevB.77.045401}. The higher polarizability
of Xe makes the dispersion interaction more stronger, which leads
to a strong binding with substrates; this inferences are inline with
the previous observation, where $E_{ads}$ increases in the increasing
sequences of atomic polarizability of rare gas atoms \citep{doi:10.1021/jp110669p}

\begin{table*}
\centering{}%
\begin{tabular}{>{\centering}m{2cm}>{\centering}m{1.5cm}>{\centering}m{4cm}>{\centering}m{4cm}}
\hline 
System & ads.site & $E_{ads}$(meV) & $d_{e}^{a}$ (\AA)\tabularnewline
\hline 
Gr -Xe & H & -175 \{-171\textsuperscript{a} -176\textsuperscript{b}\} & 3.993 \{3.840\textsuperscript{a}, 3.548\textsuperscript{b}\}\tabularnewline
Gr - Kr & H & -133 \{ -135\textsuperscript{b}\} & 3.855 \{ 3.490\textsuperscript{b}\}\tabularnewline
BDG - Xe & B & -178 \{ -169\textsuperscript{b}\} & 3.880 \{ 3.580\textsuperscript{b}\}\tabularnewline
BDG - Kr & B & -137 \{ -123\textsuperscript{b}\} & 3.824 \{ 3.500\textsuperscript{b}\}\tabularnewline
\hline 
\end{tabular}\caption{\label{tab:ads-prop_Pristine_BDG-Gr}Comparison of adsorption energy
($E_{ads})$ and equilibrium distance between adatom \& sheet ($d_{e}^{a})$.
\protect\textsuperscript{a}Reference \citep{doi:10.1021/jp110669p},
\protect\textsuperscript{b}Reference \citep{VAZHAPPILLY2017174} }
\end{table*}
 From table \ref{tab:ads-prop_Pristine_BDG-Gr}, it is clear that
the boron doping brings only an infinitesimal improvement in the $E_{ads}$
of Xe \& Kr with respect to pristine graphene. The adsorption strength
is related to the polarizability of Xe \& Kr, which in-turn is associated
with the charge transfer process. Hence, one can employ an effective
substrate capable of making a significant change in charge transfer
process and then the polarization of adatom, which can bring a considerable
improvement in adsorption parameters. This is the main motive of the
present study, and BDG-TM sheets are employed as substrates to accomplish
this task. Table \ref{tab:ads-prop-BDG-TM-FG} shows the adsorption
properties of Xe \& Kr on different BDG-TM substrates. The $E_{ads}$
increases significantly on BDG-TMs than on the pristine BDG sheet.
This clearly shows that, TM-decoration on BDG brings a substantial
enhancement in the Xe \& Kr adsorption. Among the different BDG-TM
substrates, the BDG-Cu shows the highest $E_{ads}$ = -724 \& -521
meV for Xe \& Kr, and the enhancement is $\approx$ 307 \& 280 \%
in comparison to BDG sheet. (The efficacy of different computational
schemes, such as Hubbard correction for treating the 3\textit{d} states
of TM atoms, local density approximation and many body dispersion
energy method (MBD@rsSCS) \citep{doi:10.1063/1.4865104} on adsorption
parameters has been tested and discussed in section-A \textit{SI},
Table S1 - \textit{SI}. The substantial enhancement in $E_{ads}$
is conspicuous in all the above schemes). It's noteworthy that, the
difference in $E_{ads}$ of Xe \& Kr on BDG is -41 meV. While this
difference becomes -203 and -193 meV on BDG-Cu and BDG-Ni substrates,
respectively. This substantial difference in $E_{ads}$ guarantees
that BDG-TM substrate can distinguish between Xe and Kr, a crucial
requirement in the nuclear industry. The ratio of Xe \& Kr adsorption
energies on BDG is 1.3. It increases to 1.4 on BDG-Cu substrate and
it is in same order of previously reported values for MOFs \citep{doi:10.1021/acs.jpcc.9b06961}. 

Xe \& Kr are physisorbed on the BDG-TM substrates. The physisorption
mechanism can be explained using charge transfer induced polarization
effects \citep{koch_understanding_2013}. Both Xe \& Kr are closed
shell rare gases, and hence they don't have permanent dipole moment
in gas phase. When these rare gases approach the substrates, a measurable
dipole moment is generated at substrate-adatom interfaces due to the
charge transfer process. This phenomena is explained using two mechanism
- image force effect \citep{PhysRevLett.46.842} and Pauli's repulsion
\citep{PhysRevLett.89.096104}. In either case, the rare gases become
positively charged and substrate becomes negatively charged \citep{koch_understanding_2013}.
The positive $Q_{eff}$ values on Xe \& Kr atoms confirms the charge
transfer from the adatom to substrate. The relatively small $Q_{eff}$
of Xe \& Kr (+0.005 \& +0.003) on BDG substrate signifies that, the
charge transfer is infinitesimal; hence the polarization of adatoms
are small, which does not bring any discernible improvement in the
adsorption energy with respect to pristine graphene. Meanwhile, the
$Q_{eff}$ values of Xe \& Kr are considerably larger on BDG-TM substrates
(Table \ref{tab:ads-prop-BDG-TM-FG}). This comparatively large charge
transfer would bring a considerable change in polarization of adatoms,
which results in an enhanced adsorption on BDG-TMs substrates. In
all the cases, the $E_{ads}$ of Xe is higher than that of Kr, which
is corroborated to their low $Q_{eff}$ values in comparison with
Xe. In conclusion, the TM-decoration alters the charge distribution
at substrate-adatom interface, which in-turn brings a considerable
change in the polarization and of adatom, leading to a substantial
improvement in the adsorption energies of Xe \& Kr. On different substrates,
the $E_{ads}$ of Xe \& Kr  increases in the following sequence, BDG
- Cu > BDG - Ni > BDG - Fe > BDG - Zn. The magnitude of $Q_{eff}$
of Xe \& Kr follow the same sequence of $E_{ads}$, and hence corroborated
each other. In-short, the adatoms having large $Q_{eff}$ is highly
polarized, and thus have high $E_{ads}$; vice versa. 

\begin{figure*}
\centering{}\includegraphics[scale=2.5]{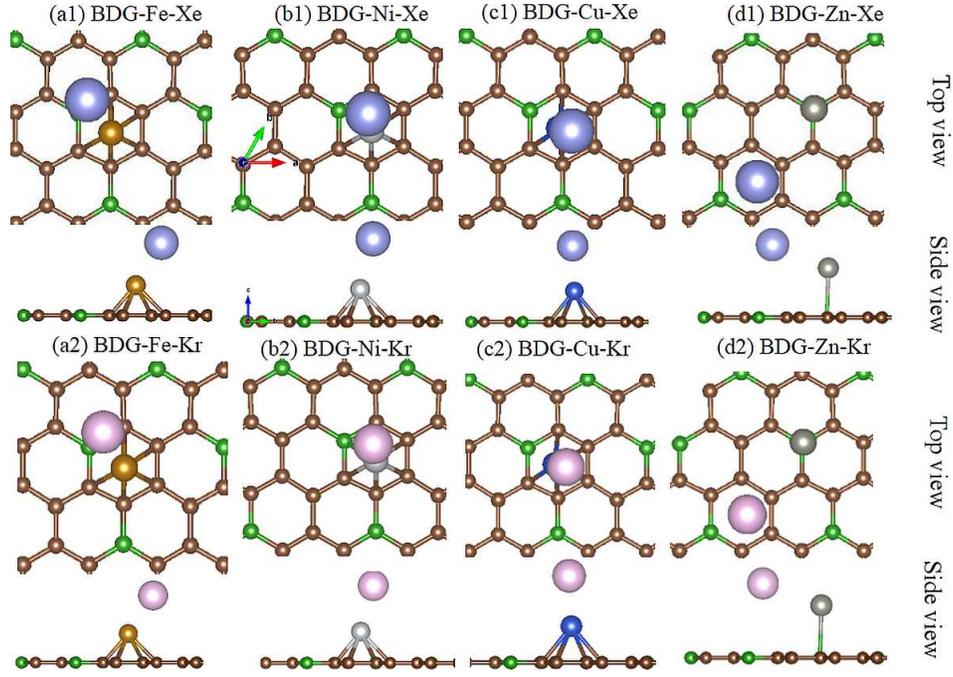}\caption{\label{fig:Optimized-geometries-Xe and Kr}Optimized geometries of
Xe \& Kr adsorbed on different BDG-TM substrates. Gold, ash, blue
and grey colour spheres represents the Fe, Ni, Cu and Zn atoms. Similarly,
violet and pink colour spheres represents, the Xe and Kr atoms, respectively. }
\end{figure*}

Figure \ref{fig:Optimized-geometries-Xe and Kr} shows the optimized
geometries of BDG-TM-adatom systems. The Xe \& Kr prefer to reside
on-top of TMs on BDG-Ni and BDG-Cu substrates, which are slightly
off-centered H site. On BDG-Fe substrate, the adatoms shift to B site,
and on BDG-Zn substrates, the adatoms drifts to a far away H site,
which is due to the strong Pauli's repulsion between the completely
filled shells of Zn and adatoms. The equilibrium distance between
the adatoms and the TMs ($d_{M}^{a}$) corroborates with $E_{ads}$.
The $d_{M}^{a}$ decreases as $E_{ads}$ increases and it's true for
both Xe \& Kr on all the BDG-TM substrates considered here. The effect
of different geometry optimization schemes on adsorption paramters
has been tested (section-B, \textit{SI}) and found to be insignificant.
Later, the thermal stability of the substrate-adatom system have been
confirmed by performing \textit{ab initio }molecular dynamics simulations
for BDG-Cu-Xe. The system is equilibrated at 300 K using velocity
re-scaling method. An integration time step of 1 \textit{fs} is adopted
and the whole simulations were done for 4000 steps. The total energy
profile remains stable during the course of simulations (Figure S3
-\textit{ SI}), indicating the finite temperature structural stability
of the system. The computed $E_{ads}$ is -710 meV, it is in the same
range of DFT predicted values (Table 2). This thermal stability analysis
contemplates the application of these novel substrates at finite temperatures.

\begin{figure*}
\centering{}\includegraphics[scale=2.5]{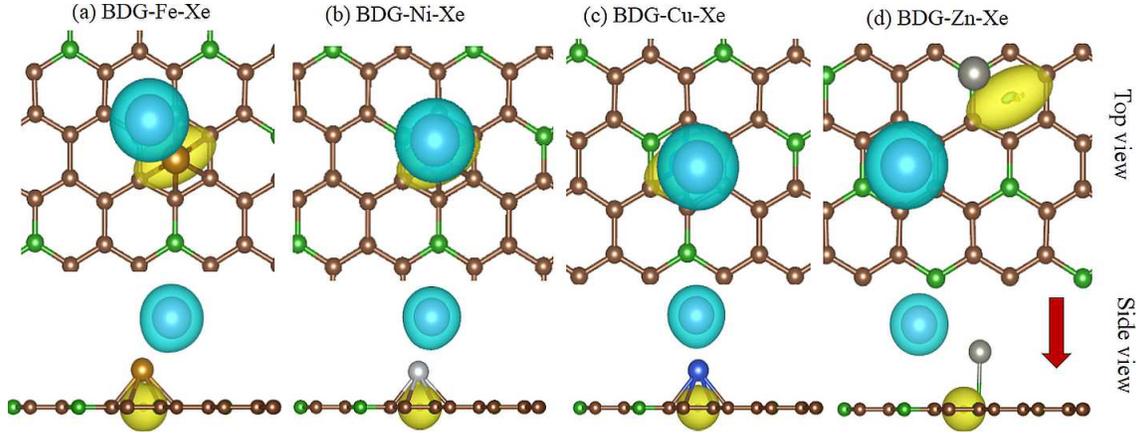}\caption{\label{fig:charge-density-differences}\textcolor{black}{Charge density
difference plots for BDG-TM-Xe; yellow and cyan region corresponds
to electron rich and deficit regions, respectively. Arrow indicates
the direction of charge transfer.}}
\end{figure*}

\begin{table*}
\centering{}%
\begin{tabular}{>{\centering}p{3cm}>{\centering}p{2cm}>{\centering}p{2cm}>{\centering}p{2cm}>{\centering}p{2cm}>{\centering}p{2.75cm}}
\toprule 
System & ads. site  & $E_{ads}$ (meV) & $d_{M}^{a}$ (\AA) & $Q_{eff}$ (TM) & $Q_{eff}$ (Xe \& Kr)\tabularnewline
\midrule
\addlinespace
BDG-Fe-Xe & B & -618 & 2.701 & +0.730 &  +0.101\tabularnewline
\addlinespace
BDG-Ni-Xe & H & -692 & 2.575 & +0.536 & +0.116\tabularnewline
\addlinespace
BDG-Cu-Xe & H & -724 & 2.558 & +0.577 &  +0.144\tabularnewline
\addlinespace
BDG-Zn-Xe & H & -212 & 3.843 & +0.341 & +0.007\tabularnewline
\addlinespace
\midrule 
BDG-Fe-Kr & B & -445 & 2.586 & +0.748 & +0.061\tabularnewline
\addlinespace
BDG-Ni-Kr & H & -499 & 2.454 & +0.576 & +0.070\tabularnewline
\addlinespace
BDG-Cu-Kr & H & -521 & 2.437 & +0.625 & +0.085\tabularnewline
\addlinespace
BDG-Zn-Kr & H & -160 & 3.815 & +0.344 & +0.004\tabularnewline
\bottomrule
\addlinespace
\end{tabular}\caption{\label{tab:ads-prop-BDG-TM-FG}Comparison of adsorption energy ($E_{ads}$),
equilibrium distance between adatom \& metal ($d_{M}^{a})$, and effective
charge ($Q_{eff})$ on transition metal (TM) \& adatoms (in units
of \textit{e}).}
\end{table*}
To get more insights on adsorption mechanism, the charge density difference
maps and partial density of states (PDOS) are analyzed. The charge
density difference maps are computed as, $\triangle\rho=\rho_{total}-\rho_{substrate}-\rho_{a}$,
where, $\rho_{total}$, $\rho_{substrate}$ and $\rho_{a}$ are the
charge densities of substrate-adatom, substrate and adatoms in gas
phase, respectively. Figure \ref{fig:charge-density-differences}
shows the charge density difference maps of Xe adsorbed on BDG-TM
substrates (Kr on BDG-TM is shown figure S4 -\textit{ SI}). Yellow
and cyan colour correspond to electron rich and deficit regions. Cyan
colour around the adatoms symbolizes that, charge transfer is from
adatoms to the BDG-TM substrates, this is true for all BDG-TM-adatom
systems considered here and is in agreement with Bader charge analysis
(Table \ref{tab:ads-prop-BDG-TM-FG}). Under polarization, there will
be a distortion of charge cloud around the adatom, which gauges the
strength of polarization. From the visual inspection (Figure \ref{fig:charge-density-differences}),
it can be seen that on BDG-Zn-adatom system, the charge cloud is almost
spherical in shape and hence distortion is least, indicating a smaller
polarization. Meanwhile, the distortion from spherical shape is conspicuous
in other BDG-TM-adatom systems and it varies as BDG-Cu-adatom > BDG-Ni-adatom
> BDG-Fe-adatom > BDG-Zn-adatom, which is a consequence of large charge
transfer and polarization effects in these system and is concomitant
with $E_{ads}$. The charge density difference maps complement the
earlier Bader charge analysis and also portrait the polarizing capability
of each substrate, which facilitating the adsorption.

\begin{figure*}
\begin{centering}
\includegraphics[scale=0.35]{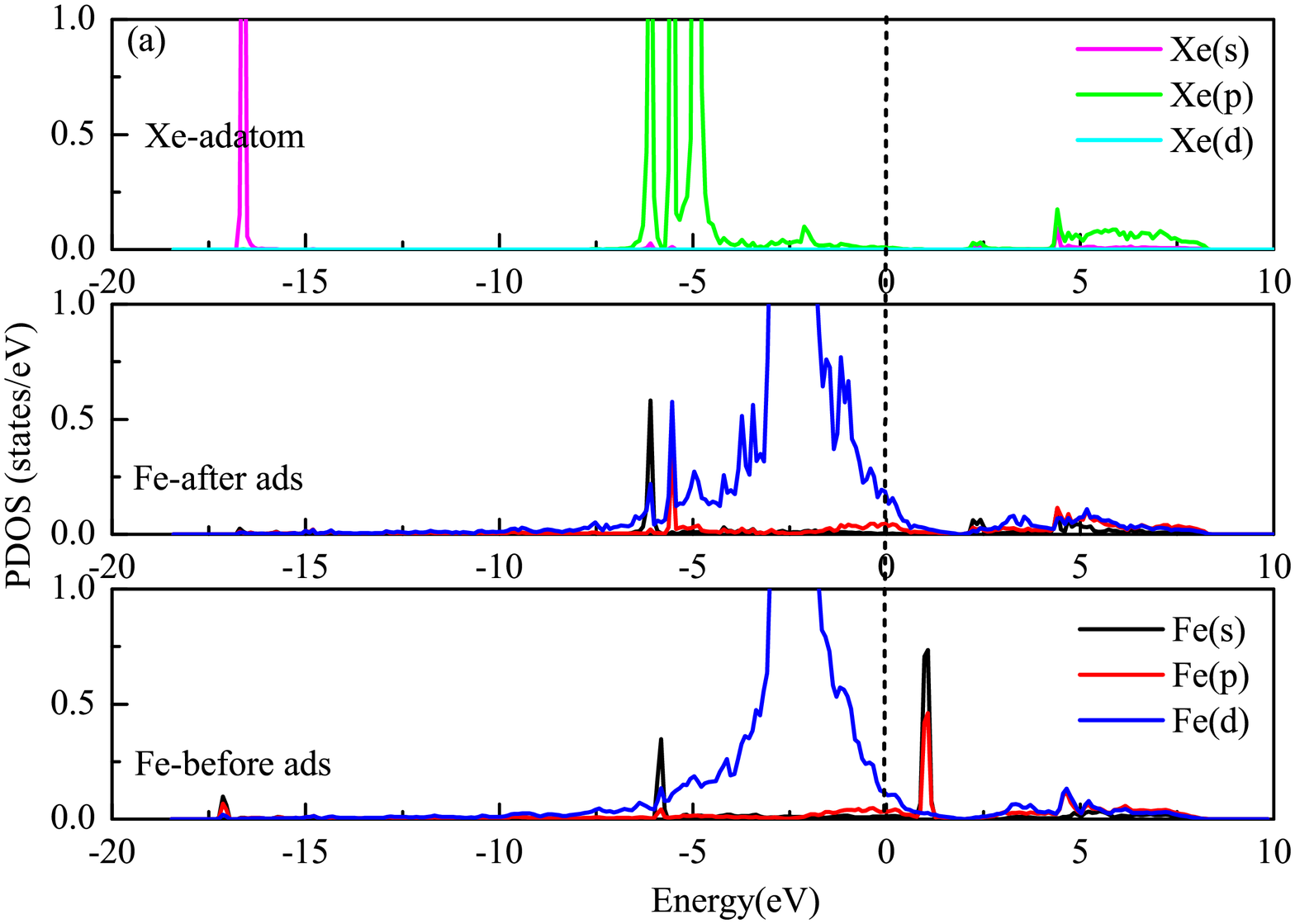}\includegraphics[scale=0.35]{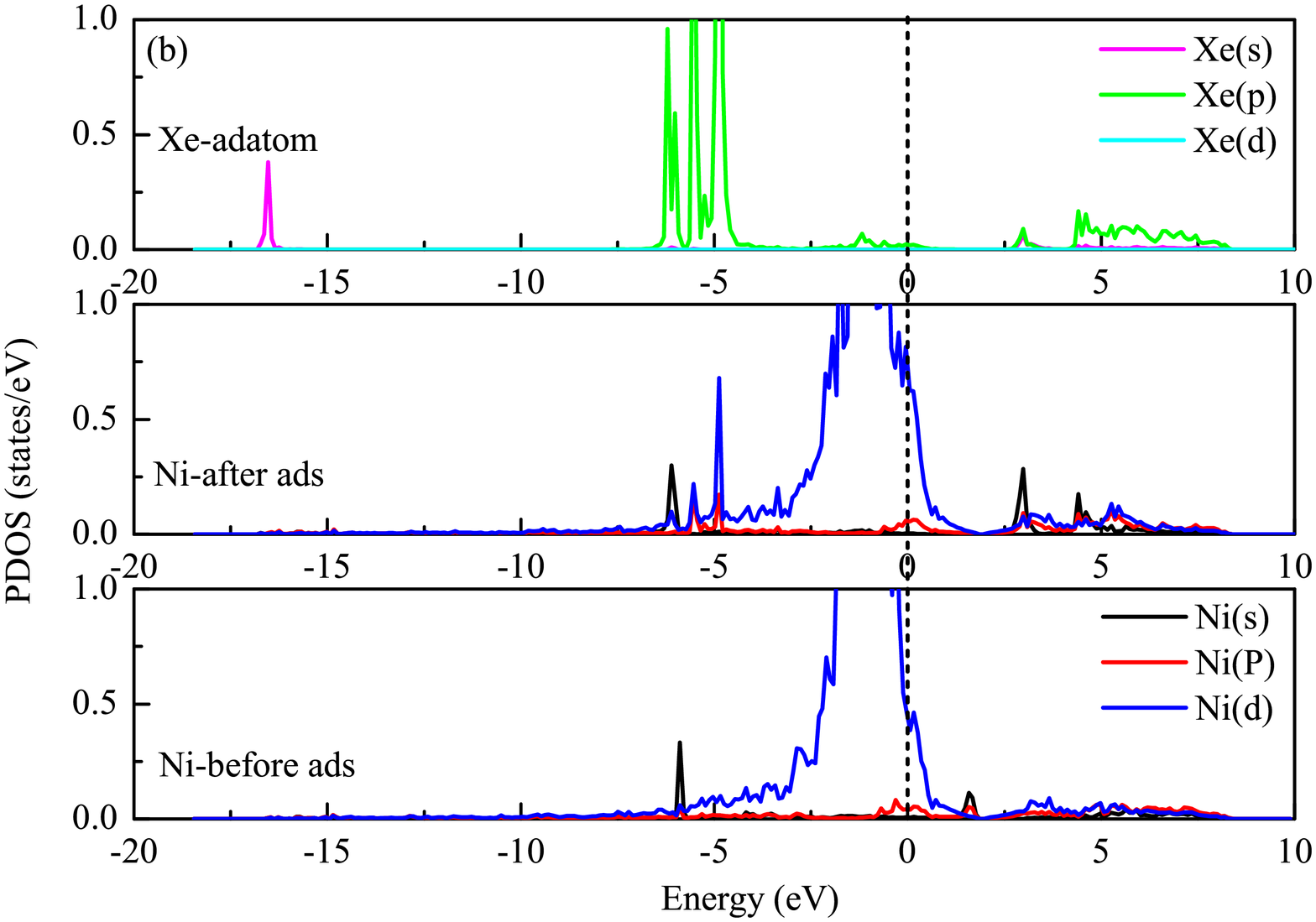}
\par\end{centering}
\centering{}\includegraphics[scale=0.35]{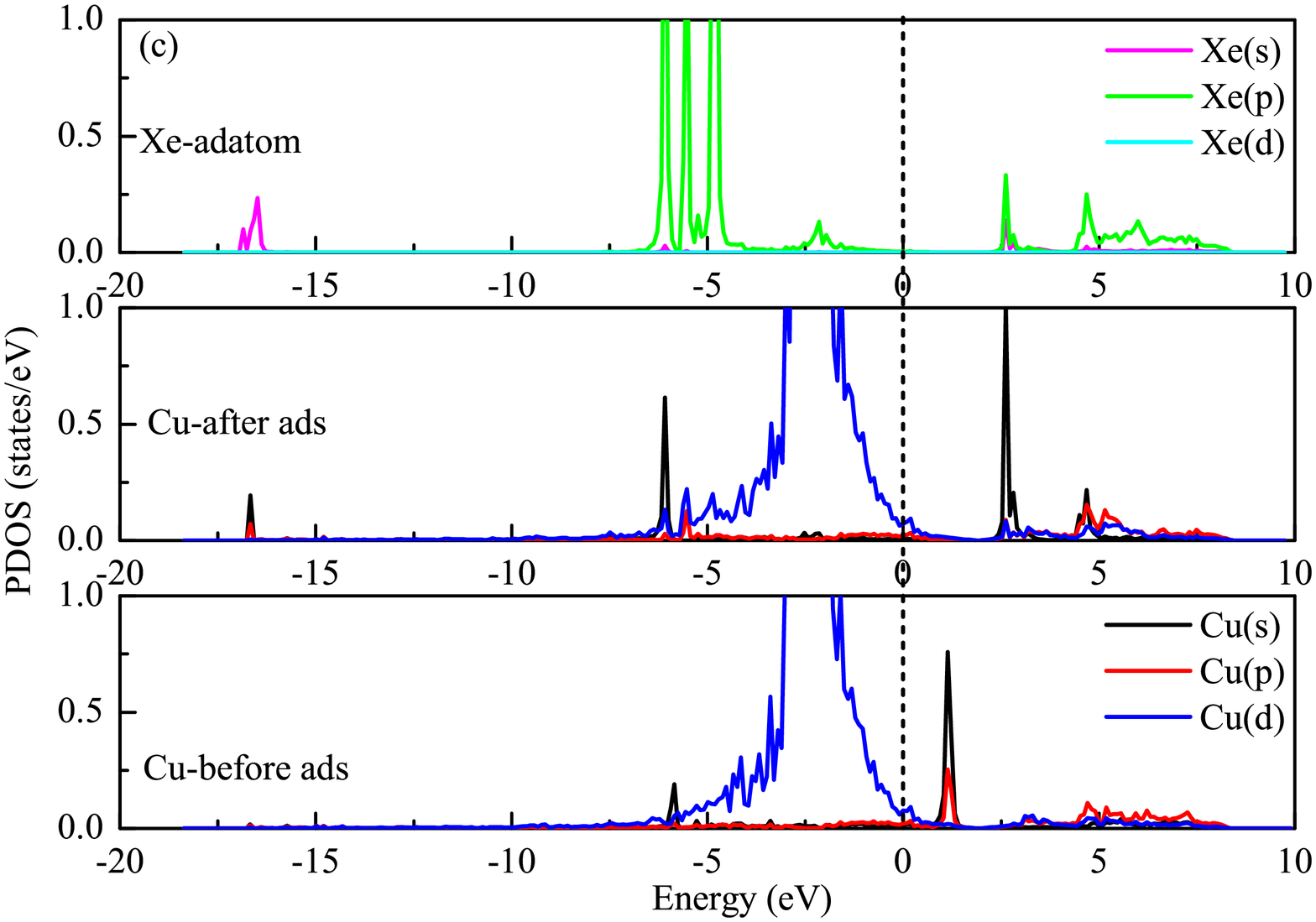}\includegraphics[scale=0.35]{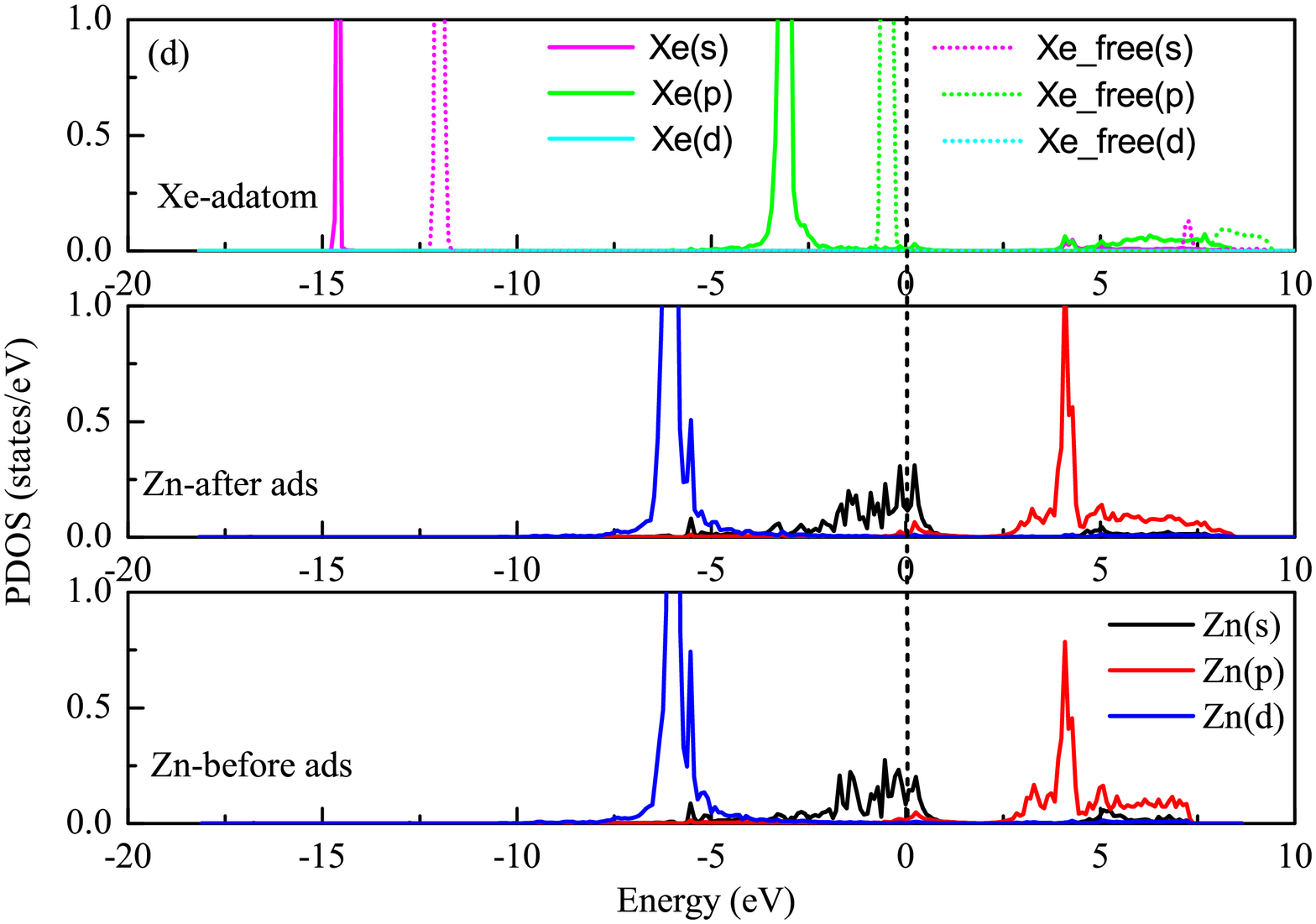}\caption{\label{fig:PDOS}Partial density of states (PDOS) of (a) BDG-Fe-Xe,
(b) BDG-Ni-Xe, (c) BDG-Cu-Xe and (d) BDG-Zn-Xe systems. In each plot,
lower, middle and top panel represents the PDOS of TM atoms before
adsorption, after adsorption and Xe adatom, respectively. PDOS of
free Xe is given in top panel of (d) as dotted line.}
\end{figure*}

Afterwards, the electronic structure of BDG-TM-adatoms are analyzed
by computing the partial density of states (PDOS). Figure \ref{fig:PDOS}
shows the PDOS of BDG-TM-Xe systems. The PDOS of free Xe is mainly
composed of \textit{s} and \textit{p} orbitals (Figure \ref{fig:PDOS}d).
Near the Fermi level, TM\textit{-d} (of Fe, Ni and cu) orbitals have
several overlapping peaks with Xe\textit{-p} orbitals, signifying
that these orbitals interact significantly upon adsorption. One can
see the wiggles on the tail (around -5 eV) of TM\textit{-d} and splitting
of Xe\textit{-p} orbitals, which are the consequences of strong interaction
of TM-\textit{d} and Xe\textit{-p} orbitals. The PDOS of BDG-Zn-Xe
(Figure \ref{fig:PDOS}d) is distinct from above systems. The Zn-\textit{d}
orbital remains intact before and after adsorption, also there is
no splitting of Xe\textit{-p} orbital upon adsorption reveals a weak
interaction of Xe-\textit{p} and Zn\textit{-d} orbital, which might
have resulted in low $E_{ads}$ on BDG-Zn substrate.

For Fe, Ni and Cu decorated system, the difference in $E_{ads}$ is
very small, hence it's difficult to have a quantitative comparison
of PDOS of these system. Silva \textit{et al} \citep{PhysRevLett.90.066104}
reported that, when the Xe adatom approaches the Pt (111) surfaces,
the Xe-\textit{p} orbitals partially depopulates and previously un-occupied
\textit{6s} and \textit{5d} states become partially occupied at closer
distances where the $E_{ads}$ is maximum. In the present case, the
depopulation of Xe-\textit{p} and partial occupancy of Xe-\textit{s}
orbital is occurring in Fe, Ni and Cu decorated systems, and it's
more apparent for BDG-Cu-Xe system (Figure \ref{fig:PDOS}c), where
the adatom-substrate equilibrium distance is minimum and $E_{ads}$
is maximum (Table \ref{tab:ads-prop-BDG-TM-FG}). Unlike metallic
substrate, adatom\textit{ d} orbital is not contributing to the PDOS
here. Below the Fermi level, the edge of Xe-\textit{p} orbitals are
distributed at -6.09, -6.20, -6.10 and -3.08 eV, for Fe, Ni, Cu and
Zn decorated system, respectively. Similarly the Xe-\textit{s} orbital
peaks at -16.60, -16.55, -16.85 and -14.59 eV for these system. Upon
adsorption, the shifting of adatom orbitals to lower energy is an
indication of charge transfer from adatom to substrate. Xe-\textit{p}
and Xe-\textit{s} orbitals shows large downshift in energy on Fe,
Ni and Cu decorated system with respect to Zn decorated, which again
substantiate the large charge transfer from Xe to these substrates.
The peak position of Xe-\textit{p} on Fe, Ni and Cu systems are close
in energy; at the same time the Xe-\textit{s} orbital of Cu-decorated
system shows appreciable downshift in energy, and also exhibits a
splitting, which might have resulted in slightly higher charge transfer
in Cu decorated system with respect to Ni \& Fe.

The change in electronic structure upon Kr adsorption is shown in
figure S5 -\textit{ SI}. Similar to BDG-TM-Xe systems, splitting of
Kr-\textit{p} and its strong interaction with TM-\textit{d} orbital
is clearly visible (around -7 eV) on Fe, Ni and Cu decorated system.
On BDG-Cu, the above effect is more conspicuous and also the Kr-\textit{p}
peak is little down in energy with respect to Fe \& Ni, leading to
a slightly more charge transfer. Unlike Xe adsorption, the depopulation
of Xe-\textit{p} and partial occupancy of Xe-\textit{s} orbital is
not apparent in Kr adsorption, which diminish the charge transfer
from Kr to substrate, leading to low polarization and $E_{ads}$.
As seen in Xe adsorption, the Kr-\textit{p} and Zn-\textit{d} orbitals
interact weakly on BDG-Zn substrate and is consistent with the low
$E_{ads}$ obtained.

\subsection{Effects of nano clustering and dispersion of TMs on adsorption properties}

\begin{figure*}
\centering{}\includegraphics[scale=2]{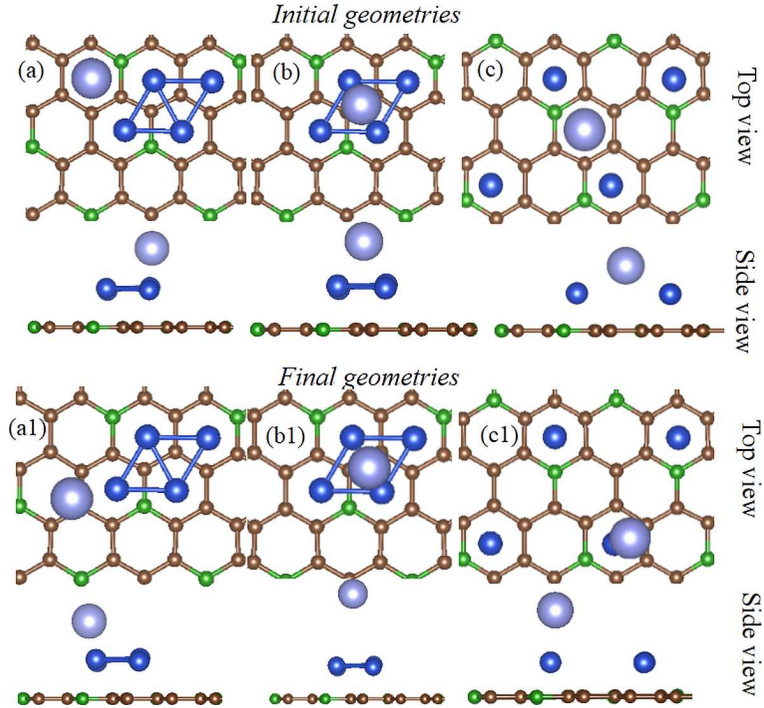}\caption{\label{fig:BDG-Cu-NC_Xe}Initial (top-panel,a-c) and final geometries
(bottom-panel, a1-c1) of; a) BDG-Cu\protect\textsubscript{4}-cluster/Xe(H-site),
b) BDG-Cu\protect\textsubscript{4}-cluster/Xe(on-top of cluster),
c) BDG-Cu-dispersed /Xe(H-site). Brown, green, blue and violet colour
spheres represent the C, B, Cu and Xe atoms, respectively. }
\end{figure*}

From experimental front, one should know the effect of clustering
and dispersion of TM atoms on $E_{ads}$ for practical applications.
In-order to analyze these effects, nano-clusters of Cu (N = 4) are
opted due to their higher adsorption capacity. Arbitrary Cu\textsubscript{4}
clusters need not guarantee a ground state configuration, hence a
genetic algorithm(GA) based technique is used to identify the global
minima (GM) \citep{doi:10.1080/23746149.2018.1516514}. To accomplish
this task, Mexican Enhanced Genetic Algorithm (MEGA) is employed;
it's a first principle based global optimization code, which combines
the genetic algorithm with parallel DFT runs to search for potential
isomers \citep{vargas_new_2017}. To obtain the GM in gas phase, 220
Cu\textsubscript{4} configurations were generated. The evolution
of gas phase Cu\textsubscript{4} clusters is shown in (Figure S6
-\textit{ SI}). The energies of 10 low-lying isomers are spread in
the range of -7.33 to -6.93 eV. Lowest energy isomer is designated
as GM and it's rhombus in shape. Subsequently, this GM structure is
placed over the BDG and relaxed to obtain the BDG-Cu\textsubscript{4}
substrate (Figure \ref{fig:BDG-Cu-NC_Xe}). To delineate the effect
of clustering and dispersion, a BDG substrate with 4-Cu atoms dispersed
at alternate H sites is considered (Figure \ref{fig:BDG-Cu-NC_Xe}c).

Further, the adsorption parameters of Xe \& Kr have been computed
on this two substrates. Here, three different cases have been considered,
a) adatom at H-site near the cluster, b) adatom on-top of the cluster,
c) adatom at H-site in Cu dispersed BDG-substrate. From table \ref{tab:Effect-of-NCs},
it's clear that the $E_{ads}$ of Xe decreases significantly on BDG
substrate with Cu\textsubscript{4} cluster in comparison to adsorption
on BDG-Cu substrate (Table \ref{tab:ads-prop-BDG-TM-FG}). In case-a,
the Xe atom moves from the initial H site to T site adjacent to cluster,
and the corresponding $E_{ads}$ = -405 meV; which is 44 \% smaller
than the value obtained on BDG-Cu substrate. When Xe atom placed on
the top of Cu\textsubscript{4} cluster (case-b), it get expelled
to a far away position from cluster, leads to very weak adsorption
(-79 meV). In the case of Cu dispersed BDG-substrate (case-c), firstly,
it can be seen that, there is no agglomeration of Cu atoms, which
is an outcome of B doping in graphene sheet \citep{BEHESHTI20111561}.
Most importantly, the $E_{ads}$ is significantly high as found earlier
(Table \ref{tab:ads-prop-BDG-TM-FG}). Clustering of Cu atoms leads
to an increase in coordination, which reduces the availability of
active binding sites, and hence become less reactive for adatoms.
This would have resulted in such drastic reduction in $E_{ads}$.
These findings are justified by the earlier observation\citep{B805179H,dobrota_chemisorption_2020};
binding energy of CO and NO on Pt cluster (Pt\textsubscript{x}) decreases
as cluster size increases. Large clusters (x > 10) become less sensitive
to CO and NO and show similar binding parameters of densely packed
Pt(111) surface. 

\begin{figure}
\centering{}\includegraphics[scale=0.4]{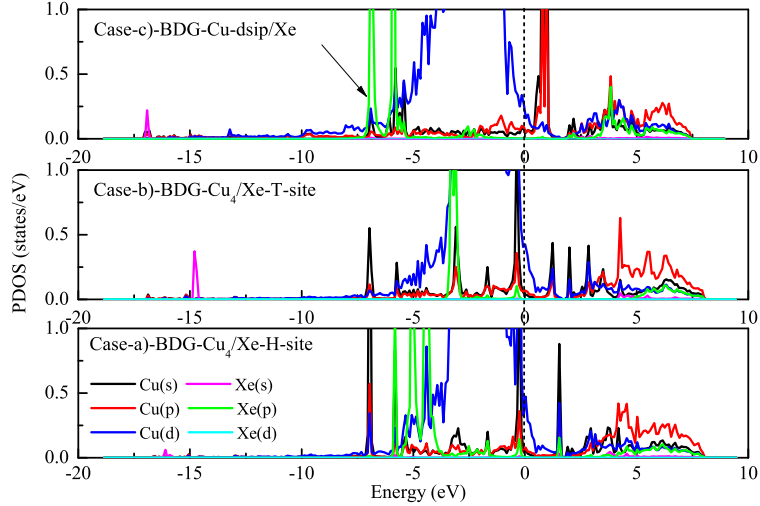}\caption{\label{fig:PDOS-CuNC-Xe}Partial density of states (PDOS) of (a) BDG-Cu\protect\textsubscript{4}/Xe-H-site,
(b) BDG-Cu\protect\textsubscript{4}/Xe-T-site, (c) BDG-Cu-disp/Xe.
Arrow shows the edge of \textit{Xe-p }peak. }
\end{figure}

Figure \ref{fig:PDOS-CuNC-Xe} shows the PDOS of above three cases.
The interaction of Xe\textit{-p }\&\textit{ }Cu\textit{-d} orbitals
is conspicuous in BDG-Cu\textsubscript{4}/Xe-H-site(case-a) and BDG-Cu-disp/Xe
systems (case-c). In BDG-Cu\textsubscript{4}/Xe-T-site (case-b),
Xe\textit{-p} \& Cu\textit{-d} orbitals interacts weakly near the
Fermi level as well as the Xe\textit{-p} edge peaks at higher energy;
which is a consequence of expulsion of Xe from the cluster. Among
case-a \& case-c, the downshift of Xe\textit{-p} and Xe\textit{-s
}orbital is comparatively higher in case-c (Figure \ref{fig:PDOS-CuNC-Xe}c),
dictating a large charge transfer and strong adsorption of adatoms
on BDG-Cu-dispersed substrate in comparison to BDG-Cu\textsubscript{4}
substrate. 

\begin{table*}
\centering{}%
\begin{tabular}{>{\centering}m{2.5cm}>{\centering}m{1cm}>{\centering}m{2cm}>{\centering}m{1.5cm}>{\centering}m{1cm}>{\centering}m{2cm}>{\centering}m{1.5cm}|>{\centering}m{1cm}>{\centering}m{2cm}>{\centering}m{1.5cm}}
\hline 
\multicolumn{7}{c}{TM- nano-clusters} & \multicolumn{3}{c}{TM-dispersion}\tabularnewline
\hline 
\hline 
System & site & $E_{ads}$(meV) & $d_{M}^{a}$ (\AA) & site & $E_{ads}$(meV) & $d_{M}^{a}$ (\AA) & site & $E_{ads}$(meV) & $d_{M}^{a}$ (\AA)\tabularnewline
\hline 
BDG -Cu\textsubscript{4}-Xe & H & -405 & 2.744 & T & -79.0 & 4.306 & H & -684 & 2.569\tabularnewline
BDG -Cu\textsubscript{4}-Kr & H & -255 & 2.697 & T & -62.0 & 4.167 & H & -487 & 2.456\tabularnewline
\hline 
\end{tabular}\caption{\label{tab:Effect-of-NCs}Effect of Cu clustering and dispersion on
adsorption parameters of Xe \& Kr.}
\end{table*}
Afterwards, the work functions (W.F) of above systems were analyzed.
The W.F of pristine BDG sheet is 5.578 eV, which is in qualitative
agreement with previous calculation \citep{C4CP01638F}. Upon metal
binding, the W.F value of BDG-Cu\textsubscript{4} cluster \& BDG
-Cu dispersed substrates are 5.332 \& 4.859 eV, respectively. Metal
binding leads to a decrease in W.F and it's more prominent in Cu dispersed
system. Table \ref{tab:Work-function} shows the W.F and change in
W.F upon adsorption of Xe \& Kr. In all the cases, the W.F value decreases
upon adsorption of Xe \& Kr. This is due to the charge transfer between
adatom and substrates. The value of W.F of substrate-adatom systems
falls below the pristine BDG, signifies that adatom posses a positive
effective charge, which have been confirmed from earlier Bader charge
analysis and charge density distribution maps. Noteworthy, the change
in W.F ($\Delta\phi=\phi_{BDG-Cu_{4}}^{Xe/Kr}-$$\phi_{BDG-Cu_{4}}$)
is more conspicuous on Cu dispersed substrate, which is corroborating
with the higher $E_{ads}$ obtained on these substrates. Later, the
induced dipole moment of above three system is quantitatively estimated
with dipole correction along the direction normal to the substrate
\citep{PhysRevB.51.4014}. From table \ref{tab:Work-function}, it
can be seen that, the induced dipole moment is highest for BDG-Cu-disp/adatom
system than the BDG-Cu\textsubscript{4 }/adatoms. The large induced
dipole field in the BDG-Cu-disp/adatom system facilitates a strong
adsorption. From the above observations, it can be concluded that,
clustering of Cu atoms deteriorate the $E_{ads}$ of Xe \& Kr. Hence
for experimental realization, uniform dispersion of fine metal particle
on BDG is preferred. In the present study, we employed a 4-atom Cu
cluster, in experiments the cluster size is in the nanometric regime.
It would be interesting to study the cluster size dependence of $E_{ads}$,
which is beyond the scope of the present work and is planned to be
carried out as an independent computational study. 

\begin{table*}
\centering{}%
\begin{tabular}{>{\centering}p{2.3cm}>{\centering}p{0.8cm}>{\centering}p{1.1cm}>{\centering}p{1.1cm}>{\centering}p{1.1cm}>{\centering}p{0.8cm}>{\centering}p{1.1cm}>{\centering}p{1.1cm}>{\centering}p{1.1cm}|>{\centering}p{0.8cm}>{\centering}p{1.1cm}>{\centering}p{1.1cm}>{\centering}p{1.1cm}}
\hline 
\multicolumn{9}{c|}{{\footnotesize{}TM- nano-clusters }} & \multicolumn{4}{c}{TM-dispersion}\tabularnewline
\hline 
{\footnotesize{}System} & {\footnotesize{}site} & {\footnotesize{}$\phi$ } & {\footnotesize{}$\Delta\phi$ } & {\footnotesize{}P } & {\footnotesize{}site} & {\footnotesize{}$\phi$} & {\footnotesize{}$\Delta\phi$ } & {\footnotesize{}P} & {\footnotesize{}site} & {\footnotesize{}$\phi$} & {\footnotesize{}$\Delta\phi$ } & {\footnotesize{}P }\tabularnewline
\hline 
{\footnotesize{}BDG -Cu}\textsubscript{{\footnotesize{}4}}{\footnotesize{}-Xe} & {\footnotesize{}H} & {\footnotesize{}5.162} & {\footnotesize{}-0.170} & {\footnotesize{}-0.815} & {\footnotesize{}T} & {\footnotesize{}5.283} & {\footnotesize{}-0.049} & {\footnotesize{}-0.535} & {\footnotesize{}H} & {\footnotesize{}4.618} & {\footnotesize{}-0.240} & {\footnotesize{}-2.435}\tabularnewline
{\footnotesize{}BDG -Cu}\textsubscript{{\footnotesize{}4}}{\footnotesize{}-Kr} & {\footnotesize{}H} & {\footnotesize{}5.225} & {\footnotesize{}-0.107} & {\footnotesize{}-0.689} & {\footnotesize{}T} & {\footnotesize{}5.302} & {\footnotesize{}-0.029} & {\footnotesize{}-0.464} & {\footnotesize{}H} & {\footnotesize{}4.676} & {\footnotesize{}-0.183} & {\footnotesize{}-2.326}\tabularnewline
\hline 
\end{tabular}\caption{\label{tab:Work-function}Work function $(\phi$) in eV, change in
work function ($\Delta\phi$) in eV and induced dipole moment (P)
in eÅ of systems }
\end{table*}
\vspace{-0.5cm}

\subsection{Comparison with prior studies}

\begin{figure*}
\centering{}\includegraphics[scale=0.5]{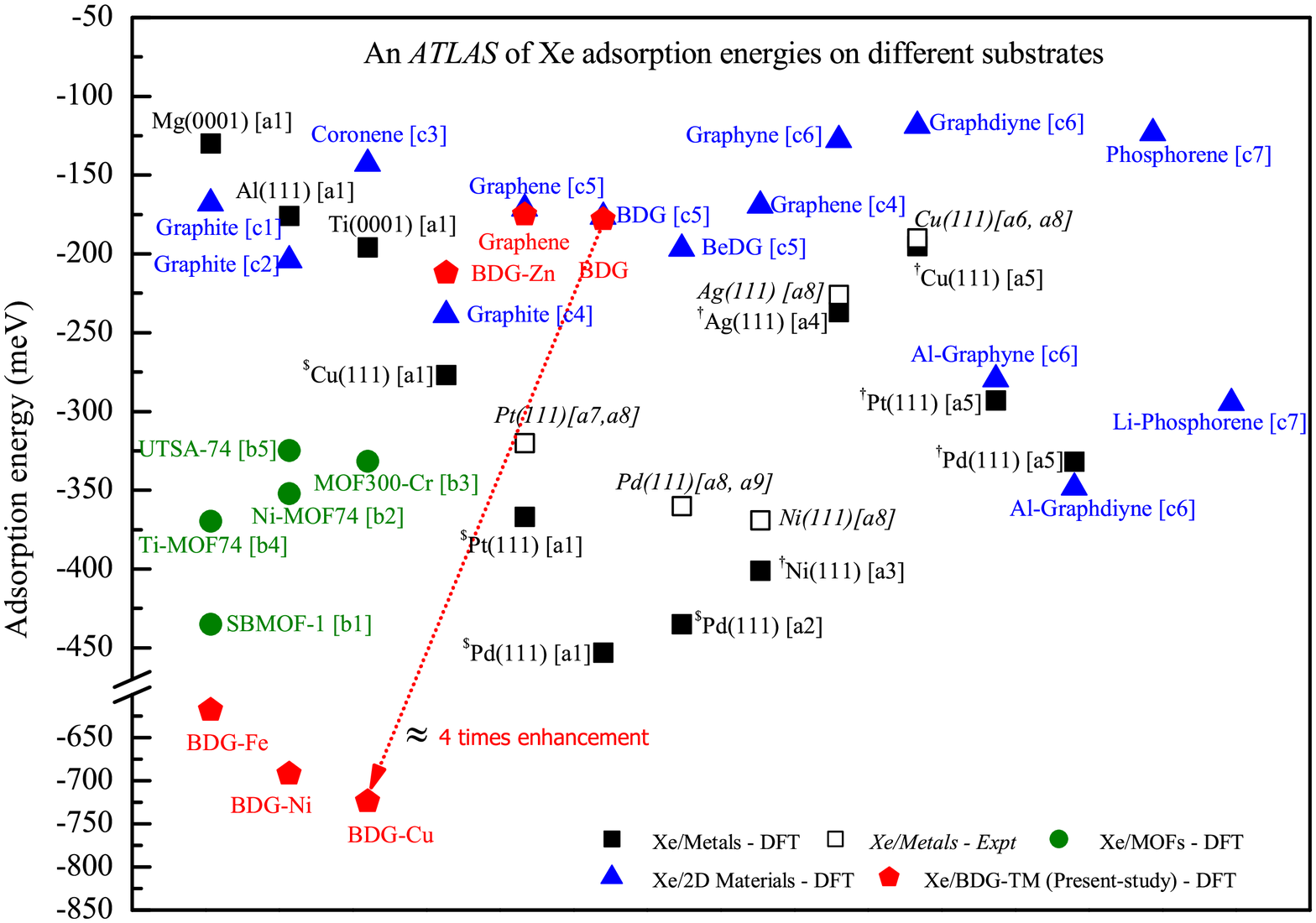}\caption{\label{fig:substrte-comp}Atlas of Xe adsorption energies ($E_{ads}$)
energies on different substrates. The filled and empty symbol represents
the computational and\textit{ experimental} data, respectively. The
selected references are, {[}a1{]} \citep{PhysRevLett.90.066104},
{[}a2{]}\citep{PhysRevB.77.045401}, {[}a3{]} \citep{PhysRevB.91.195405},
{[}a4{]} \citep{PhysRevB.93.035118}, {[}a5{]}\citep{PhysRevLett.112.106101},
{[}Expts., a6, a7, a8, a9{]}\citep{SEYLLER1998567,PhysRevB.60.11084,VIDALI1991135,PhysRevB.68.045406};
{[}b1{]}\citep{banerjee_metalorganic_2016} , {[}b2{]} \citep{doi:10.1021/acs.jpclett.5b00440},
{[}b3{]}\citep{doi:10.1021/acs.jpcc.9b06961}, {[}b4{]} \citep{doi:10.1021/acs.jpcc.6b02782}
{[}b5{]}\citep{tao_boosting_2020},; {[}c1{]}\citep{PhysRevB.76.085301},
{[}c2{]}\citep{CHEN2003403}, {[}c3{]}\citep{doi:10.1021/jp907861c},
{[}c4{]}\citep{doi:10.1021/jp110669p}, {[}c5{]} \citep{VAZHAPPILLY2017174},
{[}c6{]}\citep{VAZHAPPILLY2020100738}, {[}c7{]}\citep{HU2020144326}.
\protect\textsuperscript{\$}On metallic substrates, the LDA ( {[}a1{]}
\& {[}a2{]}) calculations over predicts the $E_{das}$ in comparison
with experiments {[}a6 - a9{]}, while the vdW corrected GGA\protect\textsuperscript{$\dagger$}calculations
{[}a3, a4, a5{]} reports the Xe adsorption satisfactorily. }
\end{figure*}

Finally, to have a quantitative comparison, an Atlas of Xe $E_{ads}$
on different substrates have been made with few representative literatures
(Figure \ref{fig:substrte-comp}). As mentioned in introduction, metallic
substrates were extensively used to adsorb the Xe atom. There is a
considerable spread in the $E_{ads}$ values of Xe on metal surfaces
among different simulations. The $E_{ads}$ of Xe/pd(111) \citep{PhysRevLett.90.066104}
system computed using local density approximation (LDA) is -453 (meV),
it's 25.8 \% higher than that of experimental value \citep{VIDALI1991135,PhysRevB.68.045406}.
This large discrepancy stems from the inherent over binding in LDA
calculations. Later, vdW corrected GGA calculations predicts the $E_{ads}$
of Xe/Pd(111) is -332 meV \citep{PhysRevLett.112.106101}, which shows
very good agreement with experimental values (-360 meV) and the deviation
is only 7.7 \% \citep{VIDALI1991135,PhysRevB.68.045406}. The similar
refinement in $E_{ads}$ of Xe/Cu(111), Xe/Ag(111), Xe/Pt(111) and
Xe/Ni(111) system is obtained with vdW correction \citep{PhysRevB.93.035118,PhysRevLett.112.106101}.
From the accurate GGA-vdW calculations, the upper bound of $E_{ads}$
on metallic substrate is fixed to -401 meV on Ni (111). 

The $E_{ads}$ on the porous structures lies in the range of -325
to -435 meV {[}green circles{]}. Noteworthy, several efficient MOFs
and porous substrates were reported in literature. Thallapally \textit{et
al} has done several remarkable works in this direction, more details
can be found in the references \citep{doi:10.1021/ar5003126,BANERJEE2018466}.
In their studies, authors quantitatively assessed the Xe adsorption
capacity on different porous materials from isosteric heat of adsorption.
A direct comparison of the isosteric heat of adsorption and DFT adsorption
energy is not meaningful. Henceforth, only selected MOF's are incorporated
in Atlas, for which the Xe $E_{ads}$ were directly calculated and
reported. Many MOFs with unsaturated open metal sites, say for example
Ni-DOBDC \citep{doi:10.1021/acs.jpclett.5b00440} is expected to perform
like BDG-TM. But the Xe $E_{ads}$ on Ni-DOBDC is in the range of
-350 meV, which is far below the BDG-TM value. This might be due to
the presence of strong screening of electrostatic fields by 2D material
substrates \citep{doi:10.1021/acs.jpclett.9b00860,AMBROSETTI2018486},
which would be helpful for not propagating the dipole fields far away
and hence facilitate the adsorption. 

2D materials with different chemical doping have the $E_{ads}$ in
the range of -118 to -348 meV {[}blue triangles{]}. It is reported
that the Xe $E_{ads}$ on Cu doped graphene is -2470 meV \citep{bin___dft_2013}.
To verify this unusually high $E_{ads}$, the calculations were repeated
and an adsorption energy of -270 meV is obtained (section-C \textit{SI},
Table S3 -\textit{ SI}). The details of calculations are given in
SI. The value obtained in the present study is in the same order of
previous reports of Xe $E_{ads}$ on doped graphene with different
metallic and non metallic impurities \citep{VAZHAPPILLY2017174},
giving qualitative assurance to the present data. On the newly employed
BDG-TM substrates, the $E_{ads}$ spans in the range of -212 to -724
meV {[}red pentagons{]}. The highest $E_{ads}$ is obtained on BDG-Cu
substrate, which is 4.06 times of the pristine BDG value. It is noteworthy
that, the enhancement on BDG-Cu substrate is approximately twice that
of the upper bounds of $E_{ads}$ on conventional metallic, MOFs and
2D material substrates (1.80, 1.66 and 2.08), respectively.

\vspace{-0.5cm}

\section{Conclusions\label{sec:Conclusions}}

In conclusion, systematic \textit{ab inito} calculations have been
performed to study the adsorption of Xe \& Kr (adatoms) on the transition-metal
atoms (TMs) decorated boron-doped graphene (BDG-TM) sheets. Substantial
enhancement in the adsorption energies ($E_{ads})$ is obtained on
BDG-TM substrates, and it varies as, BDG-Cu > BDG-Ni > BDG-Fe > BDG-Zn.
The enhancement is approximately four times that of the pristine BDG
value. TM-decoration alters the charge distribution at substrate-adatom
interface, which tinkers the polarization of adatoms, resulting in
dramatic improvement in $E_{ads}$. This is further quantified from
the effective charges on the adatoms and it follows the above sequence
of $E_{ads}$, hence self consistent with each other. Further, the
electronic density of states analysis shows a significant interaction
of TM-\textit{d} \& Xe-\textit{p} orbitals near the Fermi level in
Fe, Ni and Cu decorated system, where the enhancement is more apparent.
Afterwards, the effects of clustering and dispersion of Cu atoms on
$E_{ads}$ are analyzed, and found that clustering of Cu atoms deteriorate
the $E_{ads}$ of Xe \& Kr. Hence for experimental realization, uniform
dispersion of fine metal particle is preferred on BDG rather than
using their clusters. Finally, to have a quantitative comparison,
an Atlas of $E_{ads}$ of Xe is made. The Xe $E_{ads}$ on BDG-Cu
is approximately twice that of the upper bounds of $E_{ads}$ obtained
on conventional metallic, MOFs and 2D material substrates. The enhancement
in $E_{ads}$ on new substrates is highly encouraging and these results
pave the way for experimental realization of TM decorated BDG sheets/flakes
for efficient capture and segregation of Xe \& Kr; it may supersede
the conventional metallic and MOFs substrates in commercial as well
as in nuclear applications. 

\newpage\bibliographystyle{rsc}

\begin{mcitethebibliography}{78}
\providecommand*{\natexlab}[1]{#1}
\providecommand*{\mciteSetBstSublistMode}[1]{}
\providecommand*{\mciteSetBstMaxWidthForm}[2]{}
\providecommand*{\mciteBstWouldAddEndPuncttrue}
  {\def\EndOfBibitem{\unskip.}}
\providecommand*{\mciteBstWouldAddEndPunctfalse}
  {\let\EndOfBibitem\relax}
\providecommand*{\mciteSetBstMidEndSepPunct}[3]{}
\providecommand*{\mciteSetBstSublistLabelBeginEnd}[3]{}
\providecommand*{\EndOfBibitem}{}
\mciteSetBstSublistMode{f}
\mciteSetBstMaxWidthForm{subitem}
{(\emph{\alph{mcitesubitemcount}})}
\mciteSetBstSublistLabelBeginEnd{\mcitemaxwidthsubitemform\space}
{\relax}{\relax}

\bibitem[Soelberg \emph{et~al.}(2013)Soelberg, Garn, Greenhalgh, Law, Jubin,
  Strachan, and Thallapally]{soelberg_radioactive_2013}
N.~R. Soelberg, T.~G. Garn, M.~R. Greenhalgh, J.~D. Law, R.~Jubin, D.~M.
  Strachan and P.~K. Thallapally, \emph{Science and Technology of Nuclear
  Installations}, 2013, \textbf{2013}, 1--12\relax
\mciteBstWouldAddEndPuncttrue
\mciteSetBstMidEndSepPunct{\mcitedefaultmidpunct}
{\mcitedefaultendpunct}{\mcitedefaultseppunct}\relax
\EndOfBibitem
\bibitem[Rest \emph{et~al.}(2019)Rest, Cooper, Spino, Turnbull, Uffelen, and
  Walker]{REST2019310}
J.~Rest, M.~Cooper, J.~Spino, J.~Turnbull, P.~V. Uffelen and C.~Walker,
  \emph{Journal of Nuclear Materials}, 2019, \textbf{513}, 310 -- 345\relax
\mciteBstWouldAddEndPuncttrue
\mciteSetBstMidEndSepPunct{\mcitedefaultmidpunct}
{\mcitedefaultendpunct}{\mcitedefaultseppunct}\relax
\EndOfBibitem
\bibitem[Hoffman and Berg(2018)]{hoffman_medical_2018}
I.~Hoffman and R.~Berg, \emph{Journal of Radioanalytical and Nuclear
  Chemistry}, 2018, \textbf{318}, 165--173\relax
\mciteBstWouldAddEndPuncttrue
\mciteSetBstMidEndSepPunct{\mcitedefaultmidpunct}
{\mcitedefaultendpunct}{\mcitedefaultseppunct}\relax
\EndOfBibitem
\bibitem[Dmochowski(2009)]{Dmochowski2009}
I.~Dmochowski, \emph{Nature Chemistry}, 2009, \textbf{1}, 250--250\relax
\mciteBstWouldAddEndPuncttrue
\mciteSetBstMidEndSepPunct{\mcitedefaultmidpunct}
{\mcitedefaultendpunct}{\mcitedefaultseppunct}\relax
\EndOfBibitem
\bibitem[Bruch \emph{et~al.}(1997)Bruch, Cole, and Zaremba]{bruch1997physical}
L.~Bruch, M.~Cole and E.~Zaremba, \emph{Physical Adsorption: Forces and
  Phenomena}, Clarendon Press, 1997\relax
\mciteBstWouldAddEndPuncttrue
\mciteSetBstMidEndSepPunct{\mcitedefaultmidpunct}
{\mcitedefaultendpunct}{\mcitedefaultseppunct}\relax
\EndOfBibitem
\bibitem[Wandelt and Hulse(1984)]{doi:10.1063/1.446815}
K.~Wandelt and J.~E. Hulse, \emph{The Journal of Chemical Physics}, 1984,
  \textbf{80}, 1340--1352\relax
\mciteBstWouldAddEndPuncttrue
\mciteSetBstMidEndSepPunct{\mcitedefaultmidpunct}
{\mcitedefaultendpunct}{\mcitedefaultseppunct}\relax
\EndOfBibitem
\bibitem[Vidali \emph{et~al.}(1991)Vidali, Ihm, Kim, and Cole]{VIDALI1991135}
G.~Vidali, G.~Ihm, H.-Y. Kim and M.~W. Cole, \emph{Surface Science Reports},
  1991, \textbf{12}, 135 -- 181\relax
\mciteBstWouldAddEndPuncttrue
\mciteSetBstMidEndSepPunct{\mcitedefaultmidpunct}
{\mcitedefaultendpunct}{\mcitedefaultseppunct}\relax
\EndOfBibitem
\bibitem[Silva \emph{et~al.}(2003)Silva, Stampfl, and
  Scheffler]{PhysRevLett.90.066104}
J.~L. F.~D. Silva, C.~Stampfl and M.~Scheffler, \emph{Phys. Rev. Lett.}, 2003,
  \textbf{90}, 066104\relax
\mciteBstWouldAddEndPuncttrue
\mciteSetBstMidEndSepPunct{\mcitedefaultmidpunct}
{\mcitedefaultendpunct}{\mcitedefaultseppunct}\relax
\EndOfBibitem
\bibitem[Zhu \emph{et~al.}(2003)Zhu, Ellmer, Malissa, Brandstetter, Semrad, and
  Zeppenfeld]{PhysRevB.68.045406}
J.~Zhu, H.~Ellmer, H.~Malissa, T.~Brandstetter, D.~Semrad and P.~Zeppenfeld,
  \emph{Phys. Rev. B}, 2003, \textbf{68}, 045406\relax
\mciteBstWouldAddEndPuncttrue
\mciteSetBstMidEndSepPunct{\mcitedefaultmidpunct}
{\mcitedefaultendpunct}{\mcitedefaultseppunct}\relax
\EndOfBibitem
\bibitem[Da~Silva and Stampfl(2008)]{PhysRevB.77.045401}
J.~L.~F. Da~Silva and C.~Stampfl, \emph{Phys. Rev. B}, 2008, \textbf{77},
  045401\relax
\mciteBstWouldAddEndPuncttrue
\mciteSetBstMidEndSepPunct{\mcitedefaultmidpunct}
{\mcitedefaultendpunct}{\mcitedefaultseppunct}\relax
\EndOfBibitem
\bibitem[Chen \emph{et~al.}(2012)Chen, Al-Saidi, and Johnson]{Chen_2012}
D.-L. Chen, W.~A. Al-Saidi and J.~K. Johnson, \emph{Journal of Physics:
  Condensed Matter}, 2012, \textbf{24}, 424211\relax
\mciteBstWouldAddEndPuncttrue
\mciteSetBstMidEndSepPunct{\mcitedefaultmidpunct}
{\mcitedefaultendpunct}{\mcitedefaultseppunct}\relax
\EndOfBibitem
\bibitem[Silvestrelli and Ambrosetti(2015)]{PhysRevB.91.195405}
P.~L. Silvestrelli and A.~Ambrosetti, \emph{Phys. Rev. B}, 2015, \textbf{91},
  195405\relax
\mciteBstWouldAddEndPuncttrue
\mciteSetBstMidEndSepPunct{\mcitedefaultmidpunct}
{\mcitedefaultendpunct}{\mcitedefaultseppunct}\relax
\EndOfBibitem
\bibitem[Bazan \emph{et~al.}(2011)Bazan, Bastos-Neto, Moeller, Dreisbach, and
  Staudt]{bazan_adsorption_2011}
R.~E. Bazan, M.~Bastos-Neto, A.~Moeller, F.~Dreisbach and R.~Staudt,
  \emph{Adsorption}, 2011, \textbf{17}, 371--383\relax
\mciteBstWouldAddEndPuncttrue
\mciteSetBstMidEndSepPunct{\mcitedefaultmidpunct}
{\mcitedefaultendpunct}{\mcitedefaultseppunct}\relax
\EndOfBibitem
\bibitem[Thallapally \emph{et~al.}(2012)Thallapally, Grate, and
  Motkuri]{thallapally_facile_2012}
P.~K. Thallapally, J.~W. Grate and R.~K. Motkuri, \emph{Chem. Commun.}, 2012,
  \textbf{48}, 347--349\relax
\mciteBstWouldAddEndPuncttrue
\mciteSetBstMidEndSepPunct{\mcitedefaultmidpunct}
{\mcitedefaultendpunct}{\mcitedefaultseppunct}\relax
\EndOfBibitem
\bibitem[Banerjee \emph{et~al.}(2018)Banerjee, Simon, Elsaidi, Haranczyk, and
  Thallapally]{BANERJEE2018466}
D.~Banerjee, C.~M. Simon, S.~K. Elsaidi, M.~Haranczyk and P.~K. Thallapally,
  \emph{Chem}, 2018, \textbf{4}, 466 -- 494\relax
\mciteBstWouldAddEndPuncttrue
\mciteSetBstMidEndSepPunct{\mcitedefaultmidpunct}
{\mcitedefaultendpunct}{\mcitedefaultseppunct}\relax
\EndOfBibitem
\bibitem[Xiong \emph{et~al.}(2018)Xiong, Gong, Hu, Wu, Li, He, Chen, and
  Wang]{C7TA11321H}
S.~Xiong, Y.~Gong, S.~Hu, X.~Wu, W.~Li, Y.~He, B.~Chen and X.~Wang, \emph{J.
  Mater. Chem. A}, 2018, \textbf{6}, 4752--4758\relax
\mciteBstWouldAddEndPuncttrue
\mciteSetBstMidEndSepPunct{\mcitedefaultmidpunct}
{\mcitedefaultendpunct}{\mcitedefaultseppunct}\relax
\EndOfBibitem
\bibitem[Gong \emph{et~al.}(2018)Gong, Tang, Mao, Wu, Liu, Hu, Xiong, and
  Wang]{C8TA02091D}
Y.~Gong, Y.~Tang, Z.~Mao, X.~Wu, Q.~Liu, S.~Hu, S.~Xiong and X.~Wang, \emph{J.
  Mater. Chem. A}, 2018, \textbf{6}, 13696--13704\relax
\mciteBstWouldAddEndPuncttrue
\mciteSetBstMidEndSepPunct{\mcitedefaultmidpunct}
{\mcitedefaultendpunct}{\mcitedefaultseppunct}\relax
\EndOfBibitem
\bibitem[Wang \emph{et~al.}(2014)Wang, Yao, Zhang, Jagiello, Gong, Han, and
  Li]{C3SC52348A}
H.~Wang, K.~Yao, Z.~Zhang, J.~Jagiello, Q.~Gong, Y.~Han and J.~Li, \emph{Chem.
  Sci.}, 2014, \textbf{5}, 620--624\relax
\mciteBstWouldAddEndPuncttrue
\mciteSetBstMidEndSepPunct{\mcitedefaultmidpunct}
{\mcitedefaultendpunct}{\mcitedefaultseppunct}\relax
\EndOfBibitem
\bibitem[Banerjee \emph{et~al.}(2015)Banerjee, Cairns, Liu, Motkuri, Nune,
  Fernandez, Krishna, Strachan, and Thallapally]{doi:10.1021/ar5003126}
D.~Banerjee, A.~J. Cairns, J.~Liu, R.~K. Motkuri, S.~K. Nune, C.~A. Fernandez,
  R.~Krishna, D.~M. Strachan and P.~K. Thallapally, \emph{Accounts of Chemical
  Research}, 2015, \textbf{48}, 211--219\relax
\mciteBstWouldAddEndPuncttrue
\mciteSetBstMidEndSepPunct{\mcitedefaultmidpunct}
{\mcitedefaultendpunct}{\mcitedefaultseppunct}\relax
\EndOfBibitem
\bibitem[Perry \emph{et~al.}(2014)Perry, Teich-McGoldrick, Meek, Greathouse,
  Haranczyk, and Allendorf]{doi:10.1021/jp501495f}
J.~J. Perry, S.~L. Teich-McGoldrick, S.~T. Meek, J.~A. Greathouse, M.~Haranczyk
  and M.~D. Allendorf, \emph{The Journal of Physical Chemistry C}, 2014,
  \textbf{118}, 11685--11698\relax
\mciteBstWouldAddEndPuncttrue
\mciteSetBstMidEndSepPunct{\mcitedefaultmidpunct}
{\mcitedefaultendpunct}{\mcitedefaultseppunct}\relax
\EndOfBibitem
\bibitem[Chen \emph{et~al.}(2014)Chen, Reiss, Chong, Holden, Jelfs, Hasell,
  Little, Kewley, Briggs, Stephenson, Thomas, Armstrong, Bell, Busto, Noel,
  Liu, Strachan, Thallapally, and Cooper]{chen_separation_2014}
L.~Chen, P.~S. Reiss, S.~Y. Chong, D.~Holden, K.~E. Jelfs, T.~Hasell, M.~A.
  Little, A.~Kewley, M.~E. Briggs, A.~Stephenson, K.~M. Thomas, J.~A.
  Armstrong, J.~Bell, J.~Busto, R.~Noel, J.~Liu, D.~M. Strachan, P.~K.
  Thallapally and A.~I. Cooper, \emph{Nature Materials}, 2014, \textbf{13},
  954--960\relax
\mciteBstWouldAddEndPuncttrue
\mciteSetBstMidEndSepPunct{\mcitedefaultmidpunct}
{\mcitedefaultendpunct}{\mcitedefaultseppunct}\relax
\EndOfBibitem
\bibitem[Rahimi \emph{et~al.}(2018)Rahimi, Kamalinahad, and
  Solimannejad]{Rahimi_2018}
R.~Rahimi, S.~Kamalinahad and M.~Solimannejad, \emph{Materials Research
  Express}, 2018, \textbf{5}, 035006\relax
\mciteBstWouldAddEndPuncttrue
\mciteSetBstMidEndSepPunct{\mcitedefaultmidpunct}
{\mcitedefaultendpunct}{\mcitedefaultseppunct}\relax
\EndOfBibitem
\bibitem[Banerjee \emph{et~al.}(2016)Banerjee, Simon, Plonka, Motkuri, Liu,
  Chen, Smit, Parise, Haranczyk, and Thallapally]{banerjee_metalorganic_2016}
D.~Banerjee, C.~M. Simon, A.~M. Plonka, R.~K. Motkuri, J.~Liu, X.~Chen,
  B.~Smit, J.~B. Parise, M.~Haranczyk and P.~K. Thallapally, \emph{Nature
  Communications}, 2016, \textbf{7}, 11831\relax
\mciteBstWouldAddEndPuncttrue
\mciteSetBstMidEndSepPunct{\mcitedefaultmidpunct}
{\mcitedefaultendpunct}{\mcitedefaultseppunct}\relax
\EndOfBibitem
\bibitem[Gadipelli and Guo(2015)]{GADIPELLI20151}
S.~Gadipelli and Z.~X. Guo, \emph{Progress in Materials Science}, 2015,
  \textbf{69}, 1 -- 60\relax
\mciteBstWouldAddEndPuncttrue
\mciteSetBstMidEndSepPunct{\mcitedefaultmidpunct}
{\mcitedefaultendpunct}{\mcitedefaultseppunct}\relax
\EndOfBibitem
\bibitem[Chen \emph{et~al.}(2003)Chen, Zhou, Zhu, and Gou]{CHEN2003403}
X.-R. Chen, X.-L. Zhou, J.~Zhu and Q.-Q. Gou, \emph{Physics Letters A}, 2003,
  \textbf{315}, 403 -- 408\relax
\mciteBstWouldAddEndPuncttrue
\mciteSetBstMidEndSepPunct{\mcitedefaultmidpunct}
{\mcitedefaultendpunct}{\mcitedefaultseppunct}\relax
\EndOfBibitem
\bibitem[Pussi \emph{et~al.}(2004)Pussi, Smerdon, Ferralis, Lindroos, McGrath,
  and Diehl]{PUSSI2004157}
K.~Pussi, J.~Smerdon, N.~Ferralis, M.~Lindroos, R.~McGrath and R.~Diehl,
  \emph{Surface Science}, 2004, \textbf{548}, 157 -- 162\relax
\mciteBstWouldAddEndPuncttrue
\mciteSetBstMidEndSepPunct{\mcitedefaultmidpunct}
{\mcitedefaultendpunct}{\mcitedefaultseppunct}\relax
\EndOfBibitem
\bibitem[Da~Silva and Stampfl(2007)]{PhysRevB.76.085301}
J.~L.~F. Da~Silva and C.~Stampfl, \emph{Phys. Rev. B}, 2007, \textbf{76},
  085301\relax
\mciteBstWouldAddEndPuncttrue
\mciteSetBstMidEndSepPunct{\mcitedefaultmidpunct}
{\mcitedefaultendpunct}{\mcitedefaultseppunct}\relax
\EndOfBibitem
\bibitem[Sheng \emph{et~al.}(2010)Sheng, Ono, and
  Taketsugu]{doi:10.1021/jp907861c}
L.~Sheng, Y.~Ono and T.~Taketsugu, \emph{The Journal of Physical Chemistry C},
  2010, \textbf{114}, 3544--3548\relax
\mciteBstWouldAddEndPuncttrue
\mciteSetBstMidEndSepPunct{\mcitedefaultmidpunct}
{\mcitedefaultendpunct}{\mcitedefaultseppunct}\relax
\EndOfBibitem
\bibitem[Ambrosetti and Silvestrelli(2011)]{doi:10.1021/jp110669p}
A.~Ambrosetti and P.~L. Silvestrelli, \emph{The Journal of Physical Chemistry
  C}, 2011, \textbf{115}, 3695--3702\relax
\mciteBstWouldAddEndPuncttrue
\mciteSetBstMidEndSepPunct{\mcitedefaultmidpunct}
{\mcitedefaultendpunct}{\mcitedefaultseppunct}\relax
\EndOfBibitem
\bibitem[Vazhappilly \emph{et~al.}(2017)Vazhappilly, Ghanty, and
  Jagatap]{VAZHAPPILLY2017174}
T.~Vazhappilly, T.~K. Ghanty and B.~Jagatap, \emph{Journal of Nuclear
  Materials}, 2017, \textbf{490}, 174 -- 180\relax
\mciteBstWouldAddEndPuncttrue
\mciteSetBstMidEndSepPunct{\mcitedefaultmidpunct}
{\mcitedefaultendpunct}{\mcitedefaultseppunct}\relax
\EndOfBibitem
\bibitem[Vazhappilly and Ghanty(2020)]{VAZHAPPILLY2020100738}
T.~Vazhappilly and T.~K. Ghanty, \emph{Materials Today Communications}, 2020,
  \textbf{22}, 100738\relax
\mciteBstWouldAddEndPuncttrue
\mciteSetBstMidEndSepPunct{\mcitedefaultmidpunct}
{\mcitedefaultendpunct}{\mcitedefaultseppunct}\relax
\EndOfBibitem
\bibitem[Hu \emph{et~al.}(2020)Hu, Zhao, Du, and Jiang]{HU2020144326}
J.~Hu, L.~Zhao, J.~Du and G.~Jiang, \emph{Applied Surface Science}, 2020,
  \textbf{504}, 144326\relax
\mciteBstWouldAddEndPuncttrue
\mciteSetBstMidEndSepPunct{\mcitedefaultmidpunct}
{\mcitedefaultendpunct}{\mcitedefaultseppunct}\relax
\EndOfBibitem
\bibitem[Nachimuthu \emph{et~al.}(2014)Nachimuthu, Lai, and
  Jiang]{NACHIMUTHU2014132}
S.~Nachimuthu, P.-J. Lai and J.-C. Jiang, \emph{Carbon}, 2014, \textbf{73}, 132
  -- 140\relax
\mciteBstWouldAddEndPuncttrue
\mciteSetBstMidEndSepPunct{\mcitedefaultmidpunct}
{\mcitedefaultendpunct}{\mcitedefaultseppunct}\relax
\EndOfBibitem
\bibitem[Jungsuttiwong \emph{et~al.}(2016)Jungsuttiwong, Wongnongwa,
  Namuangruk, Kungwan, Promarak, and Kunaseth]{JUNGSUTTIWONG2016140}
S.~Jungsuttiwong, Y.~Wongnongwa, S.~Namuangruk, N.~Kungwan, V.~Promarak and
  M.~Kunaseth, \emph{Applied Surface Science}, 2016, \textbf{362}, 140 --
  145\relax
\mciteBstWouldAddEndPuncttrue
\mciteSetBstMidEndSepPunct{\mcitedefaultmidpunct}
{\mcitedefaultendpunct}{\mcitedefaultseppunct}\relax
\EndOfBibitem
\bibitem[Cort\'{e}s-Arriagada \emph{et~al.}(2018)Cort\'{e}s-Arriagada,
  Villegas-Escobar, and Ortega]{CORTESARRIAGADA2018227}
D.~Cort\'{e}s-Arriagada, N.~Villegas-Escobar and D.~E. Ortega, \emph{Applied
  Surface Science}, 2018, \textbf{427}, 227 -- 236\relax
\mciteBstWouldAddEndPuncttrue
\mciteSetBstMidEndSepPunct{\mcitedefaultmidpunct}
{\mcitedefaultendpunct}{\mcitedefaultseppunct}\relax
\EndOfBibitem
\bibitem[Sankaran \emph{et~al.}(2008)Sankaran, Viswanathan, and {Srinivasa
  Murthy}]{SANKARAN2008393}
M.~Sankaran, B.~Viswanathan and S.~{Srinivasa Murthy}, \emph{International
  Journal of Hydrogen Energy}, 2008, \textbf{33}, 393 -- 403\relax
\mciteBstWouldAddEndPuncttrue
\mciteSetBstMidEndSepPunct{\mcitedefaultmidpunct}
{\mcitedefaultendpunct}{\mcitedefaultseppunct}\relax
\EndOfBibitem
\bibitem[Wu \emph{et~al.}(2011)Wu, Fan, Kuo, and Deng]{wu_dft_2011}
H.-Y. Wu, X.~Fan, J.-L. Kuo and W.-Q. Deng, \emph{The Journal of Physical
  Chemistry C}, 2011, \textbf{115}, 9241--9249\relax
\mciteBstWouldAddEndPuncttrue
\mciteSetBstMidEndSepPunct{\mcitedefaultmidpunct}
{\mcitedefaultendpunct}{\mcitedefaultseppunct}\relax
\EndOfBibitem
\bibitem[Beheshti \emph{et~al.}(2011)Beheshti, Nojeh, and
  Servati]{BEHESHTI20111561}
E.~Beheshti, A.~Nojeh and P.~Servati, \emph{Carbon}, 2011, \textbf{49}, 1561 --
  1567\relax
\mciteBstWouldAddEndPuncttrue
\mciteSetBstMidEndSepPunct{\mcitedefaultmidpunct}
{\mcitedefaultendpunct}{\mcitedefaultseppunct}\relax
\EndOfBibitem
\bibitem[Shirasaki \emph{et~al.}(2000)Shirasaki, Derr\'{e}, \'{e}n\'{e}trier,
  Tressaud, and Flandrois]{SHIRASAKI20001461}
T.~Shirasaki, A.~Derr\'{e}, M.~M. \'{e}n\'{e}trier, A.~Tressaud and
  S.~Flandrois, \emph{Carbon}, 2000, \textbf{38}, 1461 -- 1467\relax
\mciteBstWouldAddEndPuncttrue
\mciteSetBstMidEndSepPunct{\mcitedefaultmidpunct}
{\mcitedefaultendpunct}{\mcitedefaultseppunct}\relax
\EndOfBibitem
\bibitem[Kresse and Furthm\"uller(1996)]{PhysRevB.54.11169}
G.~Kresse and J.~Furthm\"uller, \emph{Phys. Rev. B}, 1996, \textbf{54},
  11169--11186\relax
\mciteBstWouldAddEndPuncttrue
\mciteSetBstMidEndSepPunct{\mcitedefaultmidpunct}
{\mcitedefaultendpunct}{\mcitedefaultseppunct}\relax
\EndOfBibitem
\bibitem[Perdew \emph{et~al.}(1996)Perdew, Burke, and
  Ernzerhof]{PhysRevLett.77.3865}
J.~P. Perdew, K.~Burke and M.~Ernzerhof, \emph{Phys. Rev. Lett.}, 1996,
  \textbf{77}, 3865--3868\relax
\mciteBstWouldAddEndPuncttrue
\mciteSetBstMidEndSepPunct{\mcitedefaultmidpunct}
{\mcitedefaultendpunct}{\mcitedefaultseppunct}\relax
\EndOfBibitem
\bibitem[Monkhorst and Pack(1976)]{PhysRevB.13.5188}
H.~J. Monkhorst and J.~D. Pack, \emph{Phys. Rev. B}, 1976, \textbf{13},
  5188--5192\relax
\mciteBstWouldAddEndPuncttrue
\mciteSetBstMidEndSepPunct{\mcitedefaultmidpunct}
{\mcitedefaultendpunct}{\mcitedefaultseppunct}\relax
\EndOfBibitem
\bibitem[Vazhappilly \emph{et~al.}(2016)Vazhappilly, Ghanty, and
  Jagatap]{doi:10.1021/acs.jpcc.6b02782}
T.~Vazhappilly, T.~K. Ghanty and B.~N. Jagatap, \emph{The Journal of Physical
  Chemistry C}, 2016, \textbf{120}, 10968--10974\relax
\mciteBstWouldAddEndPuncttrue
\mciteSetBstMidEndSepPunct{\mcitedefaultmidpunct}
{\mcitedefaultendpunct}{\mcitedefaultseppunct}\relax
\EndOfBibitem
\bibitem[Tkatchenko \emph{et~al.}(2012)Tkatchenko, Distasio, Car, and
  Scheffler]{PhysRevLett.108.236402}
A.~Tkatchenko, R.~A. Distasio, R.~Car and M.~Scheffler, \emph{Phys. Rev.
  Lett.}, 2012, \textbf{108}, 236402\relax
\mciteBstWouldAddEndPuncttrue
\mciteSetBstMidEndSepPunct{\mcitedefaultmidpunct}
{\mcitedefaultendpunct}{\mcitedefaultseppunct}\relax
\EndOfBibitem
\bibitem[R\^ego \emph{et~al.}(2017)R\^ego, Tereshchuk, Oliveira, and
  Da~Silva]{PhysRevB.95.235422}
C.~R.~C. R\^ego, P.~Tereshchuk, L.~N. Oliveira and J.~L.~F. Da~Silva,
  \emph{Phys. Rev. B}, 2017, \textbf{95}, 235422\relax
\mciteBstWouldAddEndPuncttrue
\mciteSetBstMidEndSepPunct{\mcitedefaultmidpunct}
{\mcitedefaultendpunct}{\mcitedefaultseppunct}\relax
\EndOfBibitem
\bibitem[Al-Saidi \emph{et~al.}(2012)Al-Saidi, Voora, and
  Jordan]{al-saidi_assessment_2012}
W.~A. Al-Saidi, V.~K. Voora and K.~D. Jordan, \emph{Journal of Chemical Theory
  and Computation}, 2012, \textbf{8}, 1503--1513\relax
\mciteBstWouldAddEndPuncttrue
\mciteSetBstMidEndSepPunct{\mcitedefaultmidpunct}
{\mcitedefaultendpunct}{\mcitedefaultseppunct}\relax
\EndOfBibitem
\bibitem[Freire \emph{et~al.}(2018)Freire, Guedes-Sobrinho, Kiejna, and
  Da~Silva]{doi:10.1021/acs.jpcc.7b09749}
R.~L.~H. Freire, D.~Guedes-Sobrinho, A.~Kiejna and J.~L.~F. Da~Silva, \emph{The
  Journal of Physical Chemistry C}, 2018, \textbf{122}, 1577--1588\relax
\mciteBstWouldAddEndPuncttrue
\mciteSetBstMidEndSepPunct{\mcitedefaultmidpunct}
{\mcitedefaultendpunct}{\mcitedefaultseppunct}\relax
\EndOfBibitem
\bibitem[Naeini \emph{et~al.}(1996)Naeini, Way, Dahn, and
  Irwin]{PhysRevB.54.144}
J.~G. Naeini, B.~M. Way, J.~R. Dahn and J.~C. Irwin, \emph{Phys. Rev. B}, 1996,
  \textbf{54}, 144--151\relax
\mciteBstWouldAddEndPuncttrue
\mciteSetBstMidEndSepPunct{\mcitedefaultmidpunct}
{\mcitedefaultendpunct}{\mcitedefaultseppunct}\relax
\EndOfBibitem
\bibitem[Agnoli and Favaro(2016)]{C5TA10599D}
S.~Agnoli and M.~Favaro, \emph{J. Mater. Chem. A}, 2016, \textbf{4},
  5002--5025\relax
\mciteBstWouldAddEndPuncttrue
\mciteSetBstMidEndSepPunct{\mcitedefaultmidpunct}
{\mcitedefaultendpunct}{\mcitedefaultseppunct}\relax
\EndOfBibitem
\bibitem[M.~Dieb \emph{et~al.}(2018)M.~Dieb, Hou, and
  Tsuda]{doi:10.1063/1.5018065}
T.~M.~Dieb, Z.~Hou and K.~Tsuda, \emph{The Journal of Chemical Physics}, 2018,
  \textbf{148}, 241716\relax
\mciteBstWouldAddEndPuncttrue
\mciteSetBstMidEndSepPunct{\mcitedefaultmidpunct}
{\mcitedefaultendpunct}{\mcitedefaultseppunct}\relax
\EndOfBibitem
\bibitem[Nakada and Ishii(2011)]{NAKADA201113}
K.~Nakada and A.~Ishii, \emph{Solid State Communications}, 2011, \textbf{151},
  13 -- 16\relax
\mciteBstWouldAddEndPuncttrue
\mciteSetBstMidEndSepPunct{\mcitedefaultmidpunct}
{\mcitedefaultendpunct}{\mcitedefaultseppunct}\relax
\EndOfBibitem
\bibitem[Santos \emph{et~al.}(2010)Santos, Ayuela, and
  S{\'{a}}nchez-Portal]{Santos_2010}
E.~J.~G. Santos, A.~Ayuela and D.~S{\'{a}}nchez-Portal, \emph{New Journal of
  Physics}, 2010, \textbf{12}, 053012\relax
\mciteBstWouldAddEndPuncttrue
\mciteSetBstMidEndSepPunct{\mcitedefaultmidpunct}
{\mcitedefaultendpunct}{\mcitedefaultseppunct}\relax
\EndOfBibitem
\bibitem[Johll \emph{et~al.}(2009)Johll, Kang, and Tok]{PhysRevB.79.245416}
H.~Johll, H.~C. Kang and E.~S. Tok, \emph{Phys. Rev. B}, 2009, \textbf{79},
  245416\relax
\mciteBstWouldAddEndPuncttrue
\mciteSetBstMidEndSepPunct{\mcitedefaultmidpunct}
{\mcitedefaultendpunct}{\mcitedefaultseppunct}\relax
\EndOfBibitem
\bibitem[Yanagisawa \emph{et~al.}(2000)Yanagisawa, Tsuneda, and
  Hirao]{doi:10.1063/1.480546}
S.~Yanagisawa, T.~Tsuneda and K.~Hirao, \emph{The Journal of Chemical Physics},
  2000, \textbf{112}, 545--553\relax
\mciteBstWouldAddEndPuncttrue
\mciteSetBstMidEndSepPunct{\mcitedefaultmidpunct}
{\mcitedefaultendpunct}{\mcitedefaultseppunct}\relax
\EndOfBibitem
\bibitem[Tang \emph{et~al.}(2009)Tang, Sanville, and Henkelman]{Tang_2009}
W.~Tang, E.~Sanville and G.~Henkelman, \emph{Journal of Physics: Condensed
  Matter}, 2009, \textbf{21}, 084204\relax
\mciteBstWouldAddEndPuncttrue
\mciteSetBstMidEndSepPunct{\mcitedefaultmidpunct}
{\mcitedefaultendpunct}{\mcitedefaultseppunct}\relax
\EndOfBibitem
\bibitem[Thonhauser \emph{et~al.}(2007)Thonhauser, Cooper, Li, Puzder,
  Hyldgaard, and Langreth]{PhysRevB.76.125112}
T.~Thonhauser, V.~R. Cooper, S.~Li, A.~Puzder, P.~Hyldgaard and D.~C. Langreth,
  \emph{Phys. Rev. B}, 2007, \textbf{76}, 125112\relax
\mciteBstWouldAddEndPuncttrue
\mciteSetBstMidEndSepPunct{\mcitedefaultmidpunct}
{\mcitedefaultendpunct}{\mcitedefaultseppunct}\relax
\EndOfBibitem
\bibitem[Shepard \emph{et~al.}(2019)Shepard, Shepard, and Smeu]{SHEPARD201938}
R.~Shepard, S.~Shepard and M.~Smeu, \emph{Surface Science}, 2019, \textbf{682},
  38 -- 42\relax
\mciteBstWouldAddEndPuncttrue
\mciteSetBstMidEndSepPunct{\mcitedefaultmidpunct}
{\mcitedefaultendpunct}{\mcitedefaultseppunct}\relax
\EndOfBibitem
\bibitem[Kittel(2004)]{kittel2004introduction}
C.~Kittel, \emph{Introduction to Solid State Physics}, Wiley, 2004\relax
\mciteBstWouldAddEndPuncttrue
\mciteSetBstMidEndSepPunct{\mcitedefaultmidpunct}
{\mcitedefaultendpunct}{\mcitedefaultseppunct}\relax
\EndOfBibitem
\bibitem[Ambrosetti \emph{et~al.}(2014)Ambrosetti, Reilly, DiStasio, and
  Tkatchenko]{doi:10.1063/1.4865104}
A.~Ambrosetti, A.~M. Reilly, R.~A. DiStasio and A.~Tkatchenko, \emph{The
  Journal of Chemical Physics}, 2014, \textbf{140}, 18A508\relax
\mciteBstWouldAddEndPuncttrue
\mciteSetBstMidEndSepPunct{\mcitedefaultmidpunct}
{\mcitedefaultendpunct}{\mcitedefaultseppunct}\relax
\EndOfBibitem
\bibitem[Kancharlapalli \emph{et~al.}(2019)Kancharlapalli, Natarajan, and
  Ghanty]{doi:10.1021/acs.jpcc.9b06961}
S.~Kancharlapalli, S.~Natarajan and T.~K. Ghanty, \emph{The Journal of Physical
  Chemistry C}, 2019, \textbf{123}, 27531--27541\relax
\mciteBstWouldAddEndPuncttrue
\mciteSetBstMidEndSepPunct{\mcitedefaultmidpunct}
{\mcitedefaultendpunct}{\mcitedefaultseppunct}\relax
\EndOfBibitem
\bibitem[Kronik and Morikawa(2013)]{koch_understanding_2013}
L.~Kronik and Y.~Morikawa, in \emph{The {Molecule}-{Metal} {Interface}}, ed.
  N.~Koch, N.~Ueno and A.~T.~S. Wee, Wiley-VCH Verlag GmbH \& Co. KGaA,
  Weinheim, Germany, 2013, pp. 51--89\relax
\mciteBstWouldAddEndPuncttrue
\mciteSetBstMidEndSepPunct{\mcitedefaultmidpunct}
{\mcitedefaultendpunct}{\mcitedefaultseppunct}\relax
\EndOfBibitem
\bibitem[Lang(1981)]{PhysRevLett.46.842}
N.~D. Lang, \emph{Phys. Rev. Lett.}, 1981, \textbf{46}, 842--845\relax
\mciteBstWouldAddEndPuncttrue
\mciteSetBstMidEndSepPunct{\mcitedefaultmidpunct}
{\mcitedefaultendpunct}{\mcitedefaultseppunct}\relax
\EndOfBibitem
\bibitem[Bagus \emph{et~al.}(2002)Bagus, Staemmler, and
  W\"oll]{PhysRevLett.89.096104}
P.~S. Bagus, V.~Staemmler and C.~W\"oll, \emph{Phys. Rev. Lett.}, 2002,
  \textbf{89}, 096104\relax
\mciteBstWouldAddEndPuncttrue
\mciteSetBstMidEndSepPunct{\mcitedefaultmidpunct}
{\mcitedefaultendpunct}{\mcitedefaultseppunct}\relax
\EndOfBibitem
\bibitem[J\"{a}ger \emph{et~al.}(2018)J\"{a}ger, Sch\"{a}fer, and
  Johnston]{doi:10.1080/23746149.2018.1516514}
M.~J\"{a}ger, R.~Sch\"{a}fer and R.~L. Johnston, \emph{Advances in Physics: X},
  2018, \textbf{3}, 1516514\relax
\mciteBstWouldAddEndPuncttrue
\mciteSetBstMidEndSepPunct{\mcitedefaultmidpunct}
{\mcitedefaultendpunct}{\mcitedefaultseppunct}\relax
\EndOfBibitem
\bibitem[Vargas \emph{et~al.}(2017)Vargas, Buend\'{i}a, and
  Beltr\`{a}n]{vargas_new_2017}
J.~A. Vargas, F.~Buend\'{i}a and M.~R. Beltr\`{a}n, \emph{The Journal of
  Physical Chemistry C}, 2017, \textbf{121}, 10982--10991\relax
\mciteBstWouldAddEndPuncttrue
\mciteSetBstMidEndSepPunct{\mcitedefaultmidpunct}
{\mcitedefaultendpunct}{\mcitedefaultseppunct}\relax
\EndOfBibitem
\bibitem[Xu \emph{et~al.}(2008)Xu, Getman, Shelton, and Schneider]{B805179H}
Y.~Xu, R.~B. Getman, W.~A. Shelton and W.~F. Schneider, \emph{Phys. Chem. Chem.
  Phys.}, 2008, \textbf{10}, 6009--6018\relax
\mciteBstWouldAddEndPuncttrue
\mciteSetBstMidEndSepPunct{\mcitedefaultmidpunct}
{\mcitedefaultendpunct}{\mcitedefaultseppunct}\relax
\EndOfBibitem
\bibitem[Dobrota and Pa\u{s}ti(2020)]{dobrota_chemisorption_2020}
A.~S. Dobrota and I.~A. Pa\u{s}ti, \emph{Journal of Electrochemical Science and
  Engineering}, 2020, \textbf{10}, 141--159\relax
\mciteBstWouldAddEndPuncttrue
\mciteSetBstMidEndSepPunct{\mcitedefaultmidpunct}
{\mcitedefaultendpunct}{\mcitedefaultseppunct}\relax
\EndOfBibitem
\bibitem[Lazar \emph{et~al.}(2014)Lazar, Zbo\v{r}il, Pumera, and
  Otyepka]{C4CP01638F}
P.~Lazar, R.~Zbo\v{r}il, M.~Pumera and M.~Otyepka, \emph{Phys. Chem. Chem.
  Phys.}, 2014, \textbf{16}, 14231--14235\relax
\mciteBstWouldAddEndPuncttrue
\mciteSetBstMidEndSepPunct{\mcitedefaultmidpunct}
{\mcitedefaultendpunct}{\mcitedefaultseppunct}\relax
\EndOfBibitem
\bibitem[Makov and Payne(1995)]{PhysRevB.51.4014}
G.~Makov and M.~C. Payne, \emph{Phys. Rev. B}, 1995, \textbf{51},
  4014--4022\relax
\mciteBstWouldAddEndPuncttrue
\mciteSetBstMidEndSepPunct{\mcitedefaultmidpunct}
{\mcitedefaultendpunct}{\mcitedefaultseppunct}\relax
\EndOfBibitem
\bibitem[Tao and Rappe(2014)]{PhysRevLett.112.106101}
J.~Tao and A.~M. Rappe, \emph{Phys. Rev. Lett.}, 2014, \textbf{112},
  106101\relax
\mciteBstWouldAddEndPuncttrue
\mciteSetBstMidEndSepPunct{\mcitedefaultmidpunct}
{\mcitedefaultendpunct}{\mcitedefaultseppunct}\relax
\EndOfBibitem
\bibitem[Ruiz \emph{et~al.}(2016)Ruiz, Liu, and Tkatchenko]{PhysRevB.93.035118}
V.~G. Ruiz, W.~Liu and A.~Tkatchenko, \emph{Phys. Rev. B}, 2016, \textbf{93},
  035118\relax
\mciteBstWouldAddEndPuncttrue
\mciteSetBstMidEndSepPunct{\mcitedefaultmidpunct}
{\mcitedefaultendpunct}{\mcitedefaultseppunct}\relax
\EndOfBibitem
\bibitem[Seyller \emph{et~al.}(1998)Seyller, Caragiu, Diehl, Kaukasoina, and
  Lindroos]{SEYLLER1998567}
T.~Seyller, M.~Caragiu, R.~Diehl, P.~Kaukasoina and M.~Lindroos, \emph{Chemical
  Physics Letters}, 1998, \textbf{291}, 567 -- 572\relax
\mciteBstWouldAddEndPuncttrue
\mciteSetBstMidEndSepPunct{\mcitedefaultmidpunct}
{\mcitedefaultendpunct}{\mcitedefaultseppunct}\relax
\EndOfBibitem
\bibitem[Seyller \emph{et~al.}(1999)Seyller, Caragiu, Diehl, Kaukasoina, and
  Lindroos]{PhysRevB.60.11084}
T.~Seyller, M.~Caragiu, R.~D. Diehl, P.~Kaukasoina and M.~Lindroos, \emph{Phys.
  Rev. B}, 1999, \textbf{60}, 11084--11088\relax
\mciteBstWouldAddEndPuncttrue
\mciteSetBstMidEndSepPunct{\mcitedefaultmidpunct}
{\mcitedefaultendpunct}{\mcitedefaultseppunct}\relax
\EndOfBibitem
\bibitem[Ghose \emph{et~al.}(2015)Ghose, Li, Yakovenko, Dooryhee, Ehm, Ecker,
  Dippel, Halder, Strachan, and Thallapally]{doi:10.1021/acs.jpclett.5b00440}
S.~K. Ghose, Y.~Li, A.~Yakovenko, E.~Dooryhee, L.~Ehm, L.~E. Ecker, A.-C.
  Dippel, G.~J. Halder, D.~M. Strachan and P.~K. Thallapally, \emph{The Journal
  of Physical Chemistry Letters}, 2015, \textbf{6}, 1790--1794\relax
\mciteBstWouldAddEndPuncttrue
\mciteSetBstMidEndSepPunct{\mcitedefaultmidpunct}
{\mcitedefaultendpunct}{\mcitedefaultseppunct}\relax
\EndOfBibitem
\bibitem[Tao \emph{et~al.}(2020)Tao, Fan, Xu, Feng, Krishna, and
  Luo]{tao_boosting_2020}
Y.~Tao, Y.~Fan, Z.~Xu, X.~Feng, R.~Krishna and F.~Luo, \emph{Inorganic
  Chemistry}, 2020, \textbf{59}, 11793--11800\relax
\mciteBstWouldAddEndPuncttrue
\mciteSetBstMidEndSepPunct{\mcitedefaultmidpunct}
{\mcitedefaultendpunct}{\mcitedefaultseppunct}\relax
\EndOfBibitem
\bibitem[Ambrosetti and Silvestrelli(2019)]{doi:10.1021/acs.jpclett.9b00860}
A.~Ambrosetti and P.~L. Silvestrelli, \emph{The Journal of Physical Chemistry
  Letters}, 2019, \textbf{10}, 2044--2050\relax
\mciteBstWouldAddEndPuncttrue
\mciteSetBstMidEndSepPunct{\mcitedefaultmidpunct}
{\mcitedefaultendpunct}{\mcitedefaultseppunct}\relax
\EndOfBibitem
\bibitem[Ambrosetti and Silvestrelli(2018)]{AMBROSETTI2018486}
A.~Ambrosetti and P.~L. Silvestrelli, \emph{Carbon}, 2018, \textbf{139}, 486 --
  491\relax
\mciteBstWouldAddEndPuncttrue
\mciteSetBstMidEndSepPunct{\mcitedefaultmidpunct}
{\mcitedefaultendpunct}{\mcitedefaultseppunct}\relax
\EndOfBibitem
\bibitem[BIN \emph{et~al.}(2013)BIN, DINGZHONG, and BIBO]{bin___dft_2013}
H.~BIN, Y.~DINGZHONG and C.~BIBO, \emph{Asian Journal of Chemistry}, 2013,
  \textbf{25}, 9537--9542\relax
\mciteBstWouldAddEndPuncttrue
\mciteSetBstMidEndSepPunct{\mcitedefaultmidpunct}
{\mcitedefaultendpunct}{\mcitedefaultseppunct}\relax
\EndOfBibitem
\end{mcitethebibliography}
\providecommand*{\mcitethebibliography}{\thebibliography}
\csname @ifundefined\endcsname{endmcitethebibliography}
{\let\endmcitethebibliography\endthebibliography}{}

\end{document}